\documentclass[namedreferences]{SolarPhysics}
\usepackage[optionalrh,solaenum]{spr-sola-addons} 
\usepackage{graphicx}        
\usepackage{url}             
\usepackage{amssymb}
\usepackage[usenames,dvipsnames]{color}
\definecolor{violet}{rgb}{0.5,0,0.7}

\newcommand{\Bx}{B_{\rm x}}
\newcommand{\By}{B_{\rm y}}
\newcommand{\Bh}{B_{\rm h}}
\newcommand{\Bz}{B_{\rm z}}
\newcommand{\Jz}{J_{\rm z}}
\newcommand{\Bl}{B_{\rm los}}
\newcommand{\Bt}{B_{\rm trans}}
\newcommand{\B}{B}
\newcommand{\BB}{\mbox{\boldmath$B$}}
\newcommand{\JJ}{\mbox{\boldmath$J$}}
\newcommand{\arcsec}{^{\prime\prime}}

\begin{document}
\begin{article}
\begin{opening}

\title{Modeling and Interpreting The Effects of Spatial Resolution
	on Solar Magnetic Field Maps}

\author{K~D~\surname{Leka}$^{1}$\sep
	G.~\surname{Barnes}$^{1}$}
\runningauthor{Leka and Barnes}
\runningtitle{Spatial Resolution and Magnetic Field Maps} 
\institute{$^{1}$ NorthWest Research Associates, CoRA Division,
    3380 Mitchell Ln., Boulder, CO 80301 \\
    email:\url{leka@cora.nwra.com} email:\url{graham@cora.nwra.com}}

\begin{abstract}

Different methods for simulating the effects of spatial resolution
on magnetic field maps are compared, including those commonly used
for inter-instrument comparisons.  The investigation first uses
synthetic data, and the results are confirmed with {\it Hinode}/SpectroPolarimeter data.
Four methods are examined, one which manipulates the Stokes spectra
to simulate spatial-resolution degradation, and three ``post-facto''
methods where the magnetic field maps are manipulated directly.
Throughout, statistical comparisons of the degraded maps with the
originals serve to quantify the outcomes.  Overall, we find that areas
with inferred magnetic fill fractions close to unity may be insensitive
to optical spatial resolution; areas of sub-unity fill fractions are
very sensitive.  Trends with worsening spatial resolution can include
increased average field strength, lower total flux, and a field vector
oriented closer to the line of sight.  Further-derived quantities such as
vertical current density show variations even in areas of high average
magnetic fill-fraction.  In short, unresolved maps fail to represent the
distribution of the underlying unresolved fields, and the ``post-facto''
methods generally do not reproduce the effects of a smaller telescope
aperture.  It is argued that selecting a method in order to reconcile
disparate spatial resolution effects should depend on the goal, as
one method may better preserve the field distribution, while another
can reproduce spatial resolution degradation.  The results presented
should help direct future inter-instrument comparisons. 

\end{abstract}

\end{opening}

\section{Introduction}
\label{sec:intro}

Understanding the limits of the data used to analyze and interpret
the state of a system is a necessary part of
remote-sensing science.  For more than a century, the Zeeman effect in
magnetically-sensitive spectral lines has been used to detect and
interpret the presence and character of solar magnetic fields.
Much of solar physics research relies on interpreting magnetic field
``maps'' to investigate the physical state and dynamical evolution of
the solar plasma.  Quantities such as the magnetic field strength and
direction, its variation (gradient) with space and time, the current
density (or magnetic twist, current helicity, or shear angles, as preferred),
plasma velocity vector inferred in part from the Doppler signal of the
polarization spectra, and a variety of magnetic-related forces and torques
are all of interest.  They form the basis for our understanding of 
active region structure, large-scale field structure -- even the dynamo(s), corona,
and solar wind production.  And they are all available from these measurements
of the solar magnetic field, or are they?

With advancing capability of detector technology, modulator design
and larger photon-gathering capabilities, it has become a challenge
to reconcile the differing results from different instruments that
engage different observing schemes, using different optical layouts and
telescope sizes.

Comparison efforts between instruments and their resulting magnetic field
maps are not new.  Considerable effort has gone into comparisons between
observing programs which produce the line-of-sight component over the whole solar
disk ({\it e.g.}, \opencite{mdi_mwso_2005}; \opencite{demidovetal2008}; \opencite{demidovbalthasar2009}),
as these data products provide input to heliospheric models which are the
center of both ongoing research and real-time space-weather applications.
Line selection and spectral sampling are crucial to consider 
for comparisons when the instruments and final data products may appear quite similar
(Ulrich {\it et al.}, 2002, 2009).
A challenging task is to compare instruments whose observing approaches
are very different, as in the comparisons between the scanning-slit
Advanced Stokes Polarimeter (ASP) and the filter-based SOUP
instrument \cite{bergerlites2002}, the ASP and MDI \cite{bergerlites2003},
{\it Hinode}/SP and MDI \cite{hinode_mdi_moonetal2007}, and the ASP and the
Imaging Vector Magnetograph \cite{ivm2}.  The latter comparison attempted
to evaluate the performance of two vector-field data sources, which
means including the additional complications of the linear polarization
and its data products (the component of the field perpendicular, or transverse to, the line of sight, and its
azimuthal angle) in addition to the circular polarization and 
line-of-sight magnetic field component.  Such an effort is not new
\cite{wangetal92,varsik1995,baoetal2000,zhangetal2003}, and the effort
required has not become simpler with time.

The spatial-resolution issue is the focus here.
It has come to our attention, primarily through renewed efforts to
inter-compare the performance of different facilities (the ``Vector
Magnetic Field Comparison Group'', an ad-hoc group of which the authors
are members, that the manner in which different instrumental
resolutions are incorporated into these comparisons
can lead to erroneous results, in the direction of false confidence --
implying that there is little or no impact to the resulting data due to
spatial resolution, when we argue here that this is not the case.

Below we describe a way to model the gross effects from instrumental spatial
resolution for spectro-polarimetric data, and demonstrate how this is
required in order to avoid misleading results from {\it post facto}
re-binning (``post-facto'' here meaning ``applied after the inversion
from spectra to field'', such that it is the magnetogram itself which is 
``rebinned'').  We demonstrate, using both synthetic and real data, that
spatial resolution differences do in fact lead to different results.
On a positive note, in some cases the effects of varying spatial
resolution behave in a predictable and systematic manner that depends
on the structure of the observed solar feature, a result which can guide
the interpretation of data obtained at any given spatial resolution.

\section{Demonstration: Real Data}
\label{sec:realdata1}

\begin{figure}[t!]
\vspace{0.4cm}
\centerline{\includegraphics[width=0.70\textwidth]{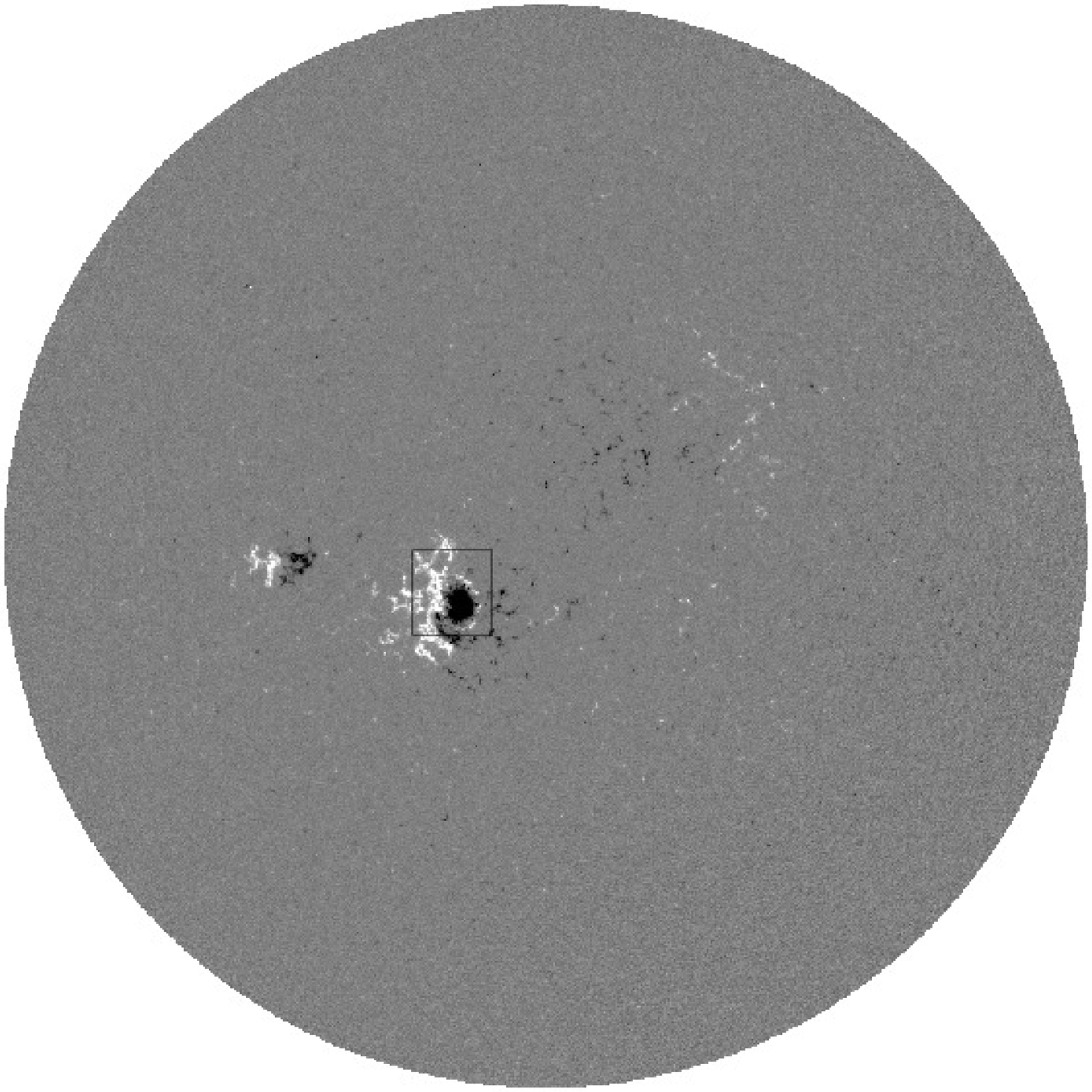}}
\vspace{-9.1cm}
\centerline{\includegraphics[width=0.3\textwidth]{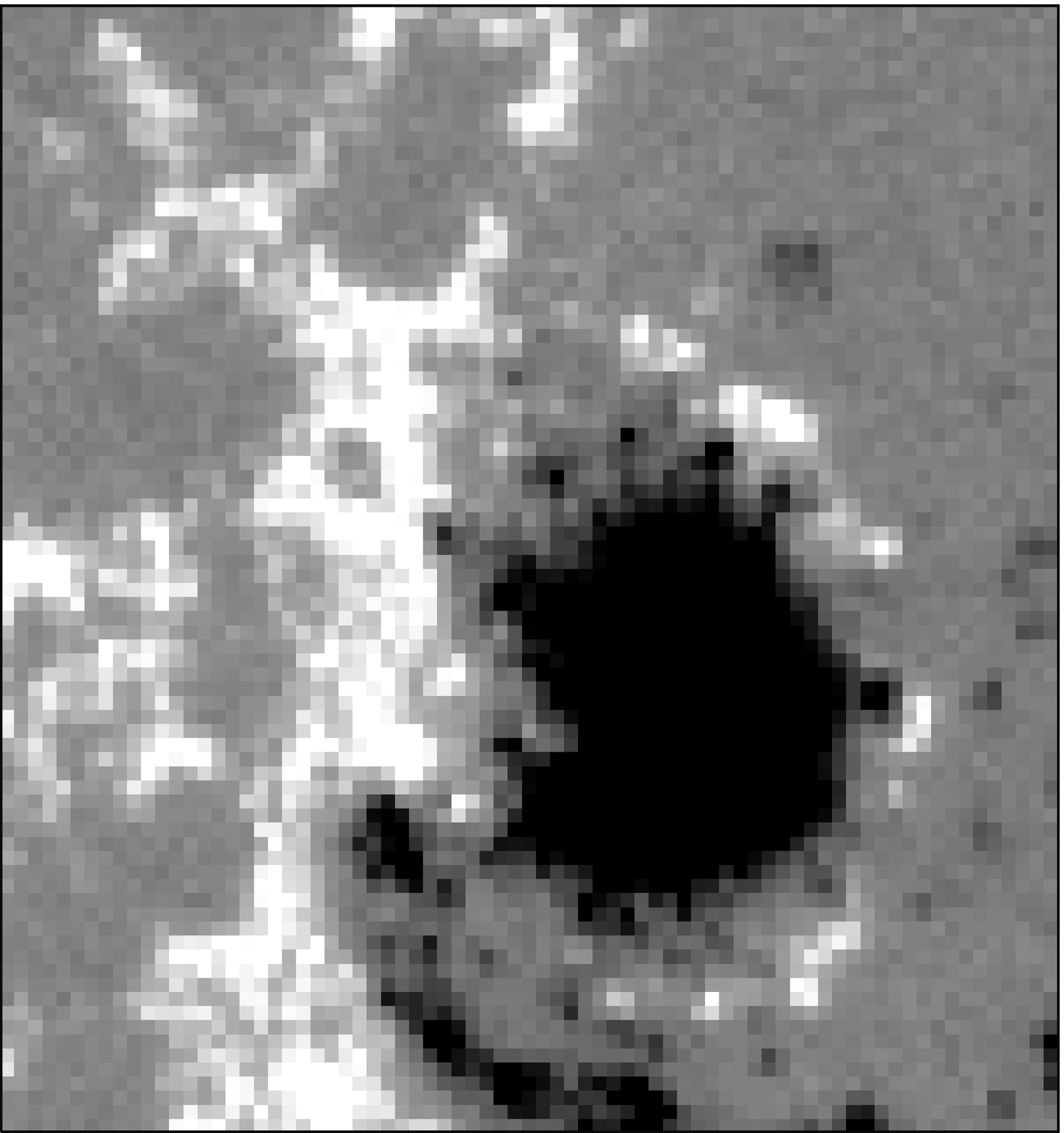}
\hspace{3cm}
\includegraphics[width=0.3\textwidth]{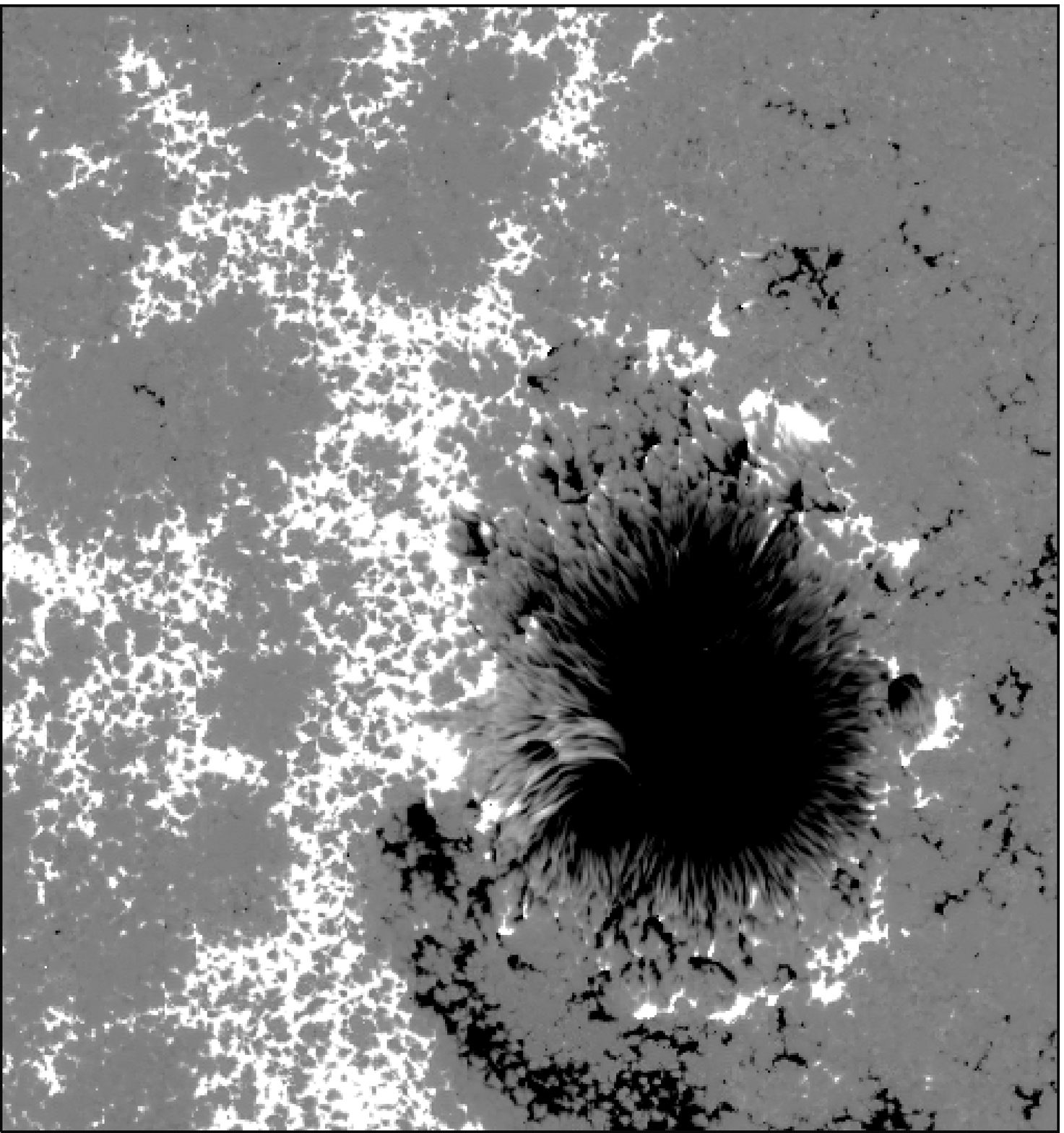}}
\vspace{5cm}
\caption{Data from the Michelson Doppler Imager (full-disk) line-of-sight
component of the ``pixel-area averaged'' field $\Bl$ at 22:24UT 30 April 2007 
includes NOAA Active Region 10953, delineated by a box.
This area is also shown magnified (left inset).  The same quantity for
the same area on the Sun, derived from a Milne-Eddington inversion of
{\it Hinode}/SpectroPolarimeter data obtained 22:30-23:15UT 30 April 2007 
is shown (right inset); all images are saturated at $\pm 500 {\rm
Mx\,cm}^{-2}$.}
\label{fig:10953blos}
\end{figure}

We begin with an example of the issue: we want to use data from two
instruments interchangeably, so how do they compare?   As an example,
we take NOAA Active Region 10953 observed on 30 April 2007.  For this
date, there exist co-temporal data from both
the Michelson Doppler Interferometer (``MDI'') aboard the {\it Solar and
Heliospheric Observatory} (``SoHO'', Scherrer {\it et al.}, 1995), and from the Solar
Optical Telescope/SpectroPolarimeter aboard the 
{\it Hinode} mission \cite{hinode,hinode_sp}; these exact data were used in
\inlinecite{nlfff3} as a boundary condition for nonlinear force-free extrapolations. The level-1.8.1 MDI ``Full-Disk Magnetogram'' from 22:24UT 30 April 2007
samples with $1.98\arcsec$ at SoHO's L-1 location, which matches the optical
spatial resolution of the telescope.  
The {\it Hinode}/SP scan which began at 22:30UT 30 April 2007\footnote{Inversion
from level-1D spectra to a magnetic map courtesy Dr.~B.W.~Lites, using the
HAO Milne-Eddington inversion code \cite{sl87} modified for {\it Hinode}/SP data, and 
presented to the authors for use in \inlinecite{nlfff3}.} is a ``fast scan'' which
performs on-chip summation for the sub-critically-sampled data, providing a final
$0.3\arcsec$-sampled map that effectively matches the telescope resolution.  
The MDI and {\it Hinode}/SP maps are shown in Figure~\ref{fig:10953blos},
where $\Bl$, the line-of-sight component of the
``pixel-area averaged'' field
is used for the {\it Hinode}/SP vector
magnetogram to ensure a consistent comparison with the MDI map, where the fill
fraction is assumed unity throughout.  (For reference, a brief table of terminology
used herein is included with Table~\ref{table:terminology}.)

\begin{table}[h!]
\begin{center}
\caption{Table of Magnetic Field Terminology.}
\label{table:terminology}
\begin{tabular}{ccc} \hline
Term & Symbol & Meaning \\
 &  (if appropriate) &  \\ \hline
Field strength & $\B$ & Magnitude of the field \\
 &  & (given in Gauss)   \\
Fill fraction & $f$  & Fraction of a pixel filled with field\\
Inclination angle & $\gamma$ & Inclination to the line of sight \\
 &  & $0^\circ,180^\circ$ along the line of sight, \\
 &  & $90^\circ$ in the plane of the sky \\
Azimuthal angle & $\phi$ & Azimuthal angle \\
``Pixel-area averaged'' & & Either $f=1.0$ is assumed, or the inferred  \\
 & & fill-fraction has been multiplied through \\ 
 & & (given in ${\rm Mx\,cm}^{-2}$). \\
Line-of-sight component & $\Bl$ &  $f \B \times \cos(\gamma)$ \\
Transverse component & $\Bt$ & $f \B \times \sin(\gamma)$ \\ \hline
\end{tabular}
\end{center}
\vspace{-1cm}
\end{table}

A sub-region of the MDI data is selected to match the {\it Hinode}/SP
field of view, to within a fraction of an MDI pixel.  The 
total of the unsigned data is computed (Table~\ref{table:mdi_sp_comp1})
at the original spatial sampling.
We then ``sampled'' the {\it Hinode}/SP $\Bl$ map using the
IDL ``{\tt congrid}'' routine
and recompute the total of the unsigned result.  No further checks
are made on the inter-instrument calibration.
We explicitly do not quote uncertainties at this point: the uncertainties
for the sums are significantly smaller than the differences between the
compared data sets, and even the effect of a bias due to 
different photon noise levels is not significant in this case.  

\begin{table}[h]
\caption{Comparison of ``Flux''.}
\label{table:mdi_sp_comp1}
\begin{tabular}{lccc} \hline
Data source & Pixel size & $\sum|B_{\rm los}| dA$ &  Difference from \\
 &  (arcsec) & ($10^{22}$ ``Mx'')  &  {\it Hinode}/SP sriginal (\%) \\ \hline
{\it Hinode}/SP & 0.317 & 2.80 &  \\
{\it Hinode}/SP & 1.98 & 2.84 & 1.2\% \\ 
MDI &  1.98 & 3.03 &  8.1\% \\ \hline
\end{tabular}
\vspace{-1cm}
\end{table}

Why is there a difference between results from {\it Hinode}/SP and MDI?  
With studies showing that MDI generally underestimates
the line-of-sight signal \cite{bergerlites2003,mdi_mwso_2005,Ulrich_etal_09},
it seems contradictory that the MDI result is the larger (see Appendix~\ref{sec:mdi_hinode_comp}).
Some difference can be attributed to the different lines
used and the different heights thus sampled ({e.g.}, \opencite{Ulrich_etal_09}),
and the different inversion methods employed.
Naively (or rhetorically) assuming that these differences are 
accounted for, the obvious remaining factor is the spatial
resolution between the two datasets.  Worse spatial resolution is
expected to dilute a polarization signal \cite{alpha2,orozco_etal_2007}; 
if this is the case, why is there only a tiny difference between the two
``resolutions'' of the {\it Hinode} data when rebinned in this manner? 

\section{Demonstration: Synthetic Data}

Light entering a polarimeter is partially polarized, with the fraction and direction 
of polarization a function of many things including the strength and 
direction of the magnetic field along the photon ray-path 
above the photospheric $\tau=1$ layer.  
Light entering a telescope
includes mixed polarization states, and optics to analyze the polarization 
generally follow the telescope entrance.
The relevant quantities regarding the effects of spatial resolution for partially
polarized light are $d$, the telescope diameter, and $I\pm P$, where $P$ is any
one (or a combination of) circular $[V]$ or linear $[Q,~U]$ polarization
signals, following the Stokes convention. 
The optical resolution varies (roughly) linearly 
with respect to $d$, meaning that the light which forms the respective
Airy disk on a resolution element (a ``pixel'') is mixed to an extent
determined by aperture size $d$ prior to analysis optics (all other elements in the system being equal).
Detected spectra are an intensity-weighted average which is a
function of $d$, meaning that bright contributions will dominate.  

\subsection{Synthesis and Treatment of Synthetic Spectra}
\label{sec:spectralbin}

To investigate and demonstrate this effect, we turn first to synthetic
data.  The approach was briefly described in \inlinecite{ambigworkshop2}, and
we present it in more detail here.  Beginning with a synthetic magnetic
model, the effects of different resolution (telescope size) on inferred
magnetic field maps are obtained as follows:

\begin{list}{$\bullet$}{
\setlength{\topsep}{0cm}   
\setlength{\partopsep}{0cm} 
\setlength{\itemsep}{0.1cm}   
\setlength{\parsep}{0cm}    
\setlength{\leftmargin}{0.5cm} 
\setlength{\rightmargin}{0cm}
\setlength{\listparindent}{0cm} 
\setlength{\itemindent}{0cm} 
\setlength{\labelsep}{0cm}   
\setlength{\labelwidth}{0cm} 
}
\item Generate emergent Stokes polarization spectra, $[I,~Q,~U,~V]$
due to the Zeeman effect on a magnetically-sensitive photospheric
line, assuming a simple Milne-Eddington atmosphere.
\item Combine the pure polarized spectra to produce ``modulated''
spectra $[I \pm P]$, {\it i.e.}, ``observed'' 
mixed-state light.
\item Manipulate these spectra as desired, along the lines of:
\begin{list}{-- }{
\setlength{\topsep}{0cm}   
\setlength{\partopsep}{0cm} 
\setlength{\itemsep}{0.1cm}   
\setlength{\parsep}{0cm}    
\setlength{\leftmargin}{0.5cm} 
\setlength{\rightmargin}{0cm}
\setlength{\listparindent}{0cm} 
\setlength{\itemindent}{0cm} 
\setlength{\labelsep}{0cm}   
\setlength{\labelwidth}{0cm} 
}
\item add simulated photon noise by drawing from a Poisson distribution 
for each particular wavelength, with the expectation value set by 
the desired ``noise level'',
\item spatially bin (by summation) the modulated spectra to a desired spatial resolution,
\item average a temporal sequence of modulated spectra from a 
target location (from a temporal sequence of synthetic maps, as appropriate),
and/or 
\item apply an instrumental response function. 
\end{list}
\item Demodulate (combine in linear combination) the manipulated spectra back to pure Stokes
$[I,~Q,~U,~V]$.
\item Re-invert using the inversion method of choice.
\end{list}

\noindent
For these tests, spectra were computed using the analytic 
Unno-Rachkovsky equations applied for the magnetic field vector and 
velocity at each pixel, and thermodynamic/line parameters typical of 
the 630.25\,nm FeI spectral line ($g_{\rm L} = 2.5$, damping $a=0.4$,
Doppler width $\lambda_{\rm D} = 0.03$\AA, absorption coefficient $\eta_0 = 10$).
Generating the Stokes spectra from the model field relied upon the spectra-genesis code which
is part of the basic Milne-Eddington least-squares inversion routine 
``{\tt stokesfit.pro}'' (available from {\it SolarSoft} distribution\footnote{\tt http://www.lmsal.com/solarsoft/ssw\_whatitis.html}).
This same inversion was then applied to the resulting Stokes
spectra to produce a magnetogram, thus the assumptions underlying
the genesis and the inversions for these test data
are internally consistent; 
the goal here is not to test inversion methods {\it per se}.
For the demonstrations here, the manipulation is limited to 
spatial binning.  

\begin{figure}
\vspace{-0.5cm}
\centerline{
\hspace{-0.50cm}
\includegraphics[width=0.58\textwidth]{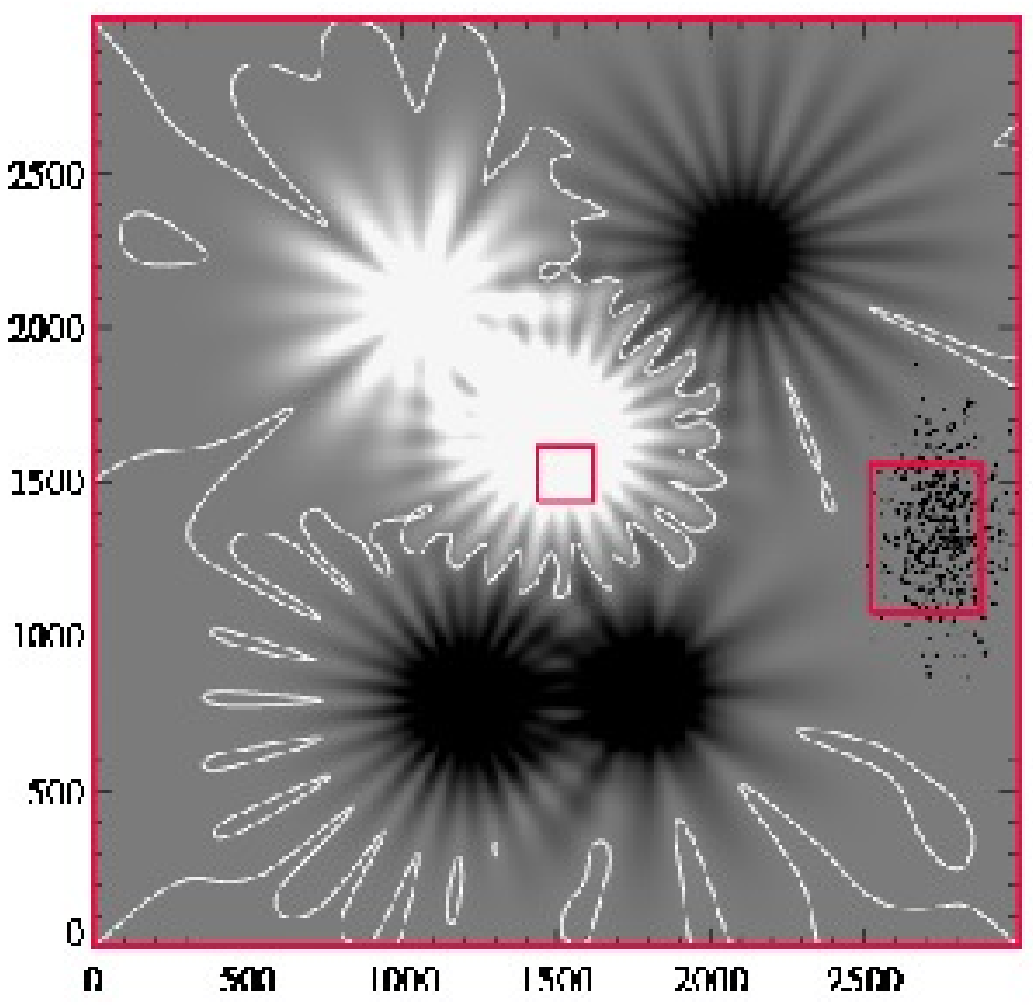}
\hspace{-1.25cm}
\includegraphics[width=0.58\textwidth]{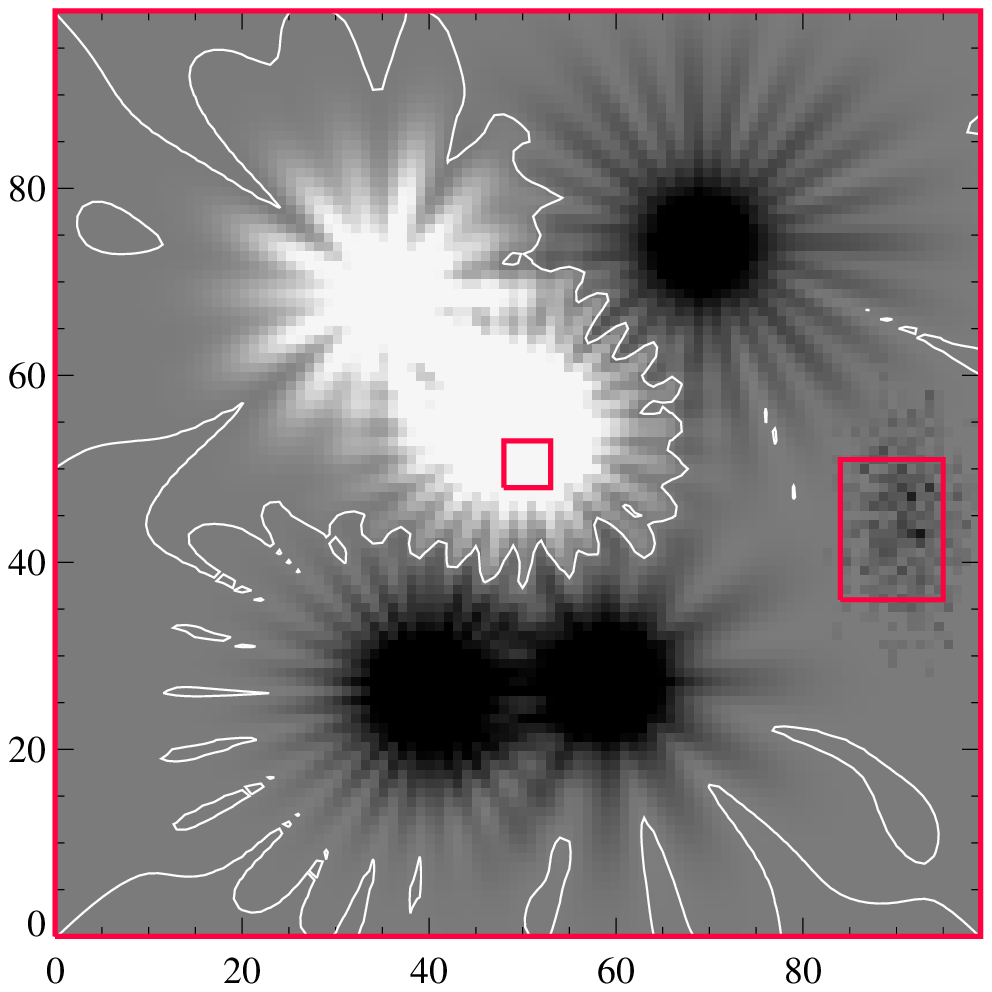}}
\vspace{-0.25cm}
\caption{\small The ``Flowers'' magnetic model
$\Bz$ component (saturating at $\pm1000$G, $\Bz > 0$ is white) 
at (left:) full resolution, $3000\times3000$ pixels arbitrarily 
set to have a $0.03\arcsec$ size.
Red boxes indicate the sub-regions highlighted in the later analysis, 
an ``umbra'' ($180\times180$ pixels) and ``plage'' ($360\times480$) areas. The smoothed
polarity inversion line is shown as a white contour.
(right:) Same, but after the spatial rebinning by a factor of 30 to a pixel 
size of $0.90\arcsec$ (using the method of spatially binning the spectra, 
see Section~\ref{sec:spectralbin}).}
\label{fig:flowers} 
\end{figure}

\subsection{The Magnetic Model}

The synthetic magnetic model has a boundary field constructed specifically to
include both areas of strong and spatially homogeneous field (reminiscent of
sunspot umbrae) and areas with significant fine-scale structure (with
features resembling penumbral fibrils and plage area).  Nicknamed the ``Flowers''
model (Figure~\ref{fig:flowers}), it is a potential-field construction
that fully satisfies Maxwell's equations.  It 
is (generally) resolved on the $3000\times3000$ computational grid,
and a $0.03\arcsec$ ``pixel size'' is assigned arbitrarily; this implies
that the magnetic fill fraction is unity for each pixel.
This synthetic boundary formed the basis of tests regarding the
effects of spatial resolution on ambiguity resolution algorithms for
vector magnetic field data \cite{ambigworkshop2}.  We refer readers to
that paper for a detailed description of its construction.  

\subsection{Signal Mixing in Spatially-Averaged Stokes Spectra}

The manipulations outlined above are the minimal steps necessary to
model the effects of an observing system.  Obviously we are completely
ignoring the details of a full optical system or spatial
smearing due to instrument jitter or atmospheric seeing effects.  In addition,
in this extremely limited demonstration we are completely ignoring any
substantive difference between an imaging system and a slit-spectrograph
polarimeter, and we are ignoring photon noise.  Of additional
note: there are no velocities in this synthetic model, which simplifies
the spectral-mixing effects considerably: no asymmetries or additional
broadening is introduced to the spectra.  In short, the present
study uses the simplest possible case.

\begin{figure}[t]
\centerline{
\hspace{-0.25cm}
\includegraphics[width=0.45\textwidth]{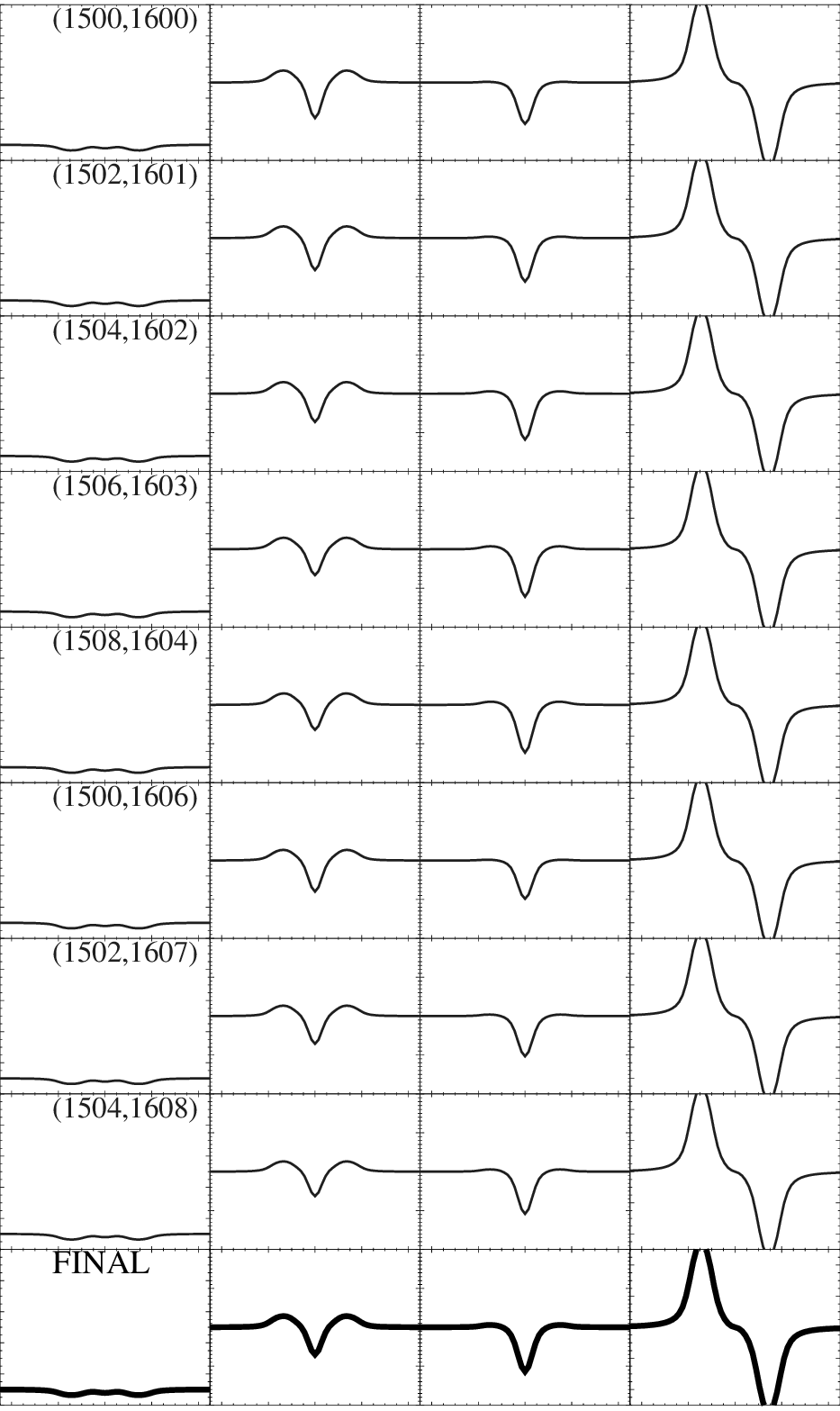}
\hspace{0.25cm}
\includegraphics[width=0.45\textwidth]{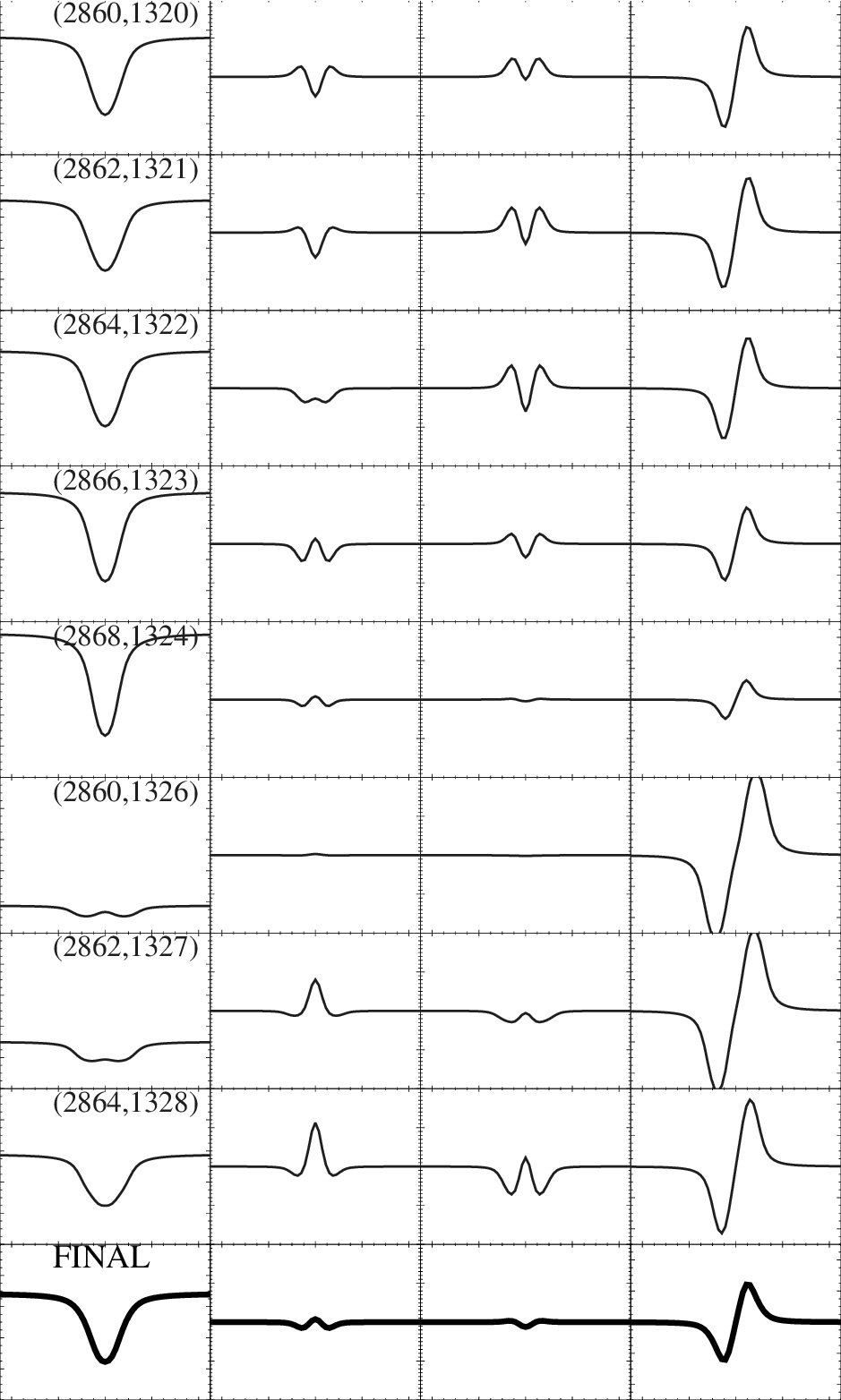}}
\caption{\small Left column: Eight samples of emergent Stokes
$[I,~Q,~U,~V]$ spectra, from a small patch (10$\times$10 pixels) of
the original synthetic (fully resolved) data, centered in an ``umbra'' at [1505,~1605]
in Figure~\ref{fig:flowers} (left).  Stokes $[I,~Q,~U,~V]$ are plotted
left-right with ranges: $I:[0,1],~Q,~U:[-0.2,0.2],~V:[-0.5,0.5]$, the
pixel coordinates (of the original model) are also shown.  Left,
bottom: the resulting ``FINAL'' $[I,~Q,~U,~V]$ after averaging the
100 underlying emergent polarization spectra, plotted on the same scale.
For this case, the resulting average is very similar to any of the sample
contributing spectra.  Right column: Same as left set, but for a
10$\times$10 pixel area centered on the ``plage'' area, at [2865,~1325]
in Figure~\ref{fig:flowers} (left).  In this case, the variability of the
underlying spectra (top) leads to an average which differs noticeably
from that arising from any single contributing pixel.}
\label{fig:spectra}
\end{figure}

We perform the spatial binning for a wide range of factors ranging
from 2 to 60.  We also include a unity bin factor, in order to have a
consistent treatment of the spectra/inversion for comparison, rather
than comparing to the raw synthetic model; in practice (as discussed in
\opencite{ambigworkshop2}) only a few pixels of the 9 million in the bin-1
case differ by more than machine precision from the original model field.

The effects of spatial resolution on the detected spectra are demonstrated
in Figure~\ref{fig:spectra}.  Consider two $10\times 10$-pixel portions
of the boundary, centered in the ``umbra'' and in the ``plage'',
respectively.  For each, samples from the 100 emergent demodulated
Stokes spectra are shown.  The emergent spectra for the umbral area
are spatially very consistent (Figure~\ref{fig:spectra} left), and
the results of averaging the underlying 100 spectra are very similar
to any individual contributing emergent spectra.  On the contrary, the
emergent spectra from the plage area (Figure~\ref{fig:spectra} right)
is spatially quite variable.  There results a significant difference
between the ``spatially binned'' resulting Stokes spectra and any single
emergent spectrum from the underlying area.

Limited resolution causes an intensity-weighted averaging of the emergent
Stokes polarization signals.  It is often clear (from multiple lobes and
extreme asymmetries, see \opencite{mismas_early}; \opencite{sigetal99};
\opencite{ugdsss00}) that the resulting observed spectra are inconsistent
with a single magnetic field vector in a simple atmosphere having a
linear source function and no additional gradients of any sort within the
resolution element (the Milne-Eddington Unno-Rachkovsky assumptions).
But sometimes it is not so clear \cite{mismas}.  Since the underlying
brightness distribution is unknown, untangling the weighting of the
contributing spectra is impossible.  This quick demonstration clearly
cautions that while a strong signal cannot be created from nothing
(instrumental and seeing effects aside, as well as any Doppler effects),
a small or nonexistent signal can result even when there are strong
underlying fields.

\subsection{Creating Magnetograms}

We now test the effects of the spatial binning of the polarization spectra on
the ability of an inversion algorithm to retrieve the underlying structure.
The synthetic binned spectra underwent an inversion using 
``{\tt stokesfit.pro}''\footnote{Implementation details: $[I,~Q,~U,~V]$ default relative weighting: $1/[10,2,2,1]$,
fill-fraction is fit, the initial guess set to the spatially binned parameters
from the original model ({\it i.e.} as close to the solution as possible),
``{\tt curvefit}'' specified unless a bad fit returned, in which case ``{\tt amoeba}''
and ``{\tt genetic}'' algorithms invoked for optimization.} which solves for 
the magnitude of the field in the instrument-frame $\Bx^i,~\By^i,~\Bz^i$,
and separately the magnetic fill fraction $f$ (see Table~\ref{table:terminology}).
The resulting magnetograms were then ambiguity
resolved using the minimum-energy code ``ME0''\footnote{available at
\url{http://www.cora.nwra.com/AMBIG/}}, described in \inlinecite{ambigworkshop2}, 
\inlinecite{hinode2ambig}.
All parameters used for the inversion and ME0 were the same for each
resolution (except those that scaled with array size), 
as it is not the intent to test either the inversion
or the ambiguity resolution algorithms {\it per se}.  What results are
vector magnetic field maps that simulate what would be observed through
telescopes when solely the aperture size varies.

\begin{table}[h]
\begin{center}
\caption{Summary and Specifics of Binning Approaches.}
\label{table:binning}
\begin{tabular}{lccl} \hline
Moniker & Algorithm & Code used & Details \\ \hline
{\bf ``instrument''} & Average modulated & ``{\tt awnoise.pro}''${}^1$ & ``bin-5'' implies  \\
& Stokes spectra & (modified)  & averaging $5\times5$ \\ 
& & & spectra, then inverting \\ \hline
\multicolumn{4}{|c|}{``Post-Facto'' Approaches} \\ \hline
{\bf \color{red} ``average''} & Simple average & IDL ``{\tt rebin}'' & Acts on image-plane \\
& $B^i_{\rm new} = N_{\rm bin}^{-2} \sum_{j=1}^{N_{\rm bin}^2} B^i_j$, & {\tt sample=0} & field components${}^2$ and \\ 
& & & field strength,  \\ 
& & & fill fraction \\ \hline
{\bf \color{green} ``bicubic''} & Bicubic Interpolation & ``{\tt brebin.pro}''${}^1$ & Acts on ambiguity- \\
& with $\JJ\times\BB = 0$ & & resolved magnetograms \\ \hline
{\bf \color{violet} ``sampled''} & Simple sampling & IDL ``{\tt congrid}'' & If bin is odd: \\
& of image-plane field & {\tt center=1}, & use center point \\
& components${}^2$ & {\tt interpolate=1}, & If bin is even: \\
& and field strength, & & use average of central  \\
& fill fraction & & four points \\  \hline
\end{tabular}
\end{center}
\vspace{-0.5cm}
\small{${}^1$ Available as part of \\
{\tt http://www.cora.nwra.com/AMBIGUITY\_WORKSHOP/2005/CODES/mgram.tar} \\ 
${}^2$ Image-plane field components are defined as $\Bx^i = \Bt\cos(\phi)$, $\By^i = \Bt\sin(\phi)$ 
$\Bz^i = \Bl$, and are used to avoid wrap at $\phi=0,2\pi$.  
}
\vspace{-0.5cm}
\end{table}

For comparison we perform three types of {\it ``post facto''} binning on the bin-1 
synthetic magnetogram, as summarized in Table~\ref{table:binning}.
The three utilize a simple averaging (referred to as {\bf \color{red} ``average''}),
a more sophisticated interpolation method developed by Dr.~T.~Metcalf
specifically for the task of sampling vector magnetograms ({\bf \color{green} ``interpolate''}),
and a sampling approach which performs a minimal amount of averaging ({\bf \color{violet} ``sampled''}).
We use color here and throughout for reference and clarity as the results of 
these methods are compared.  For each of the ``post facto'' approaches, the 
azimuthal ambiguity resolution is an acute-angle method, matched to the results from ME0 
for the ``instrumental'' approach at the same binning factor.

\subsection{Comparing the Magnetograms}

\begin{figure}
\centerline{
\includegraphics[width=0.57\textwidth,clip=]{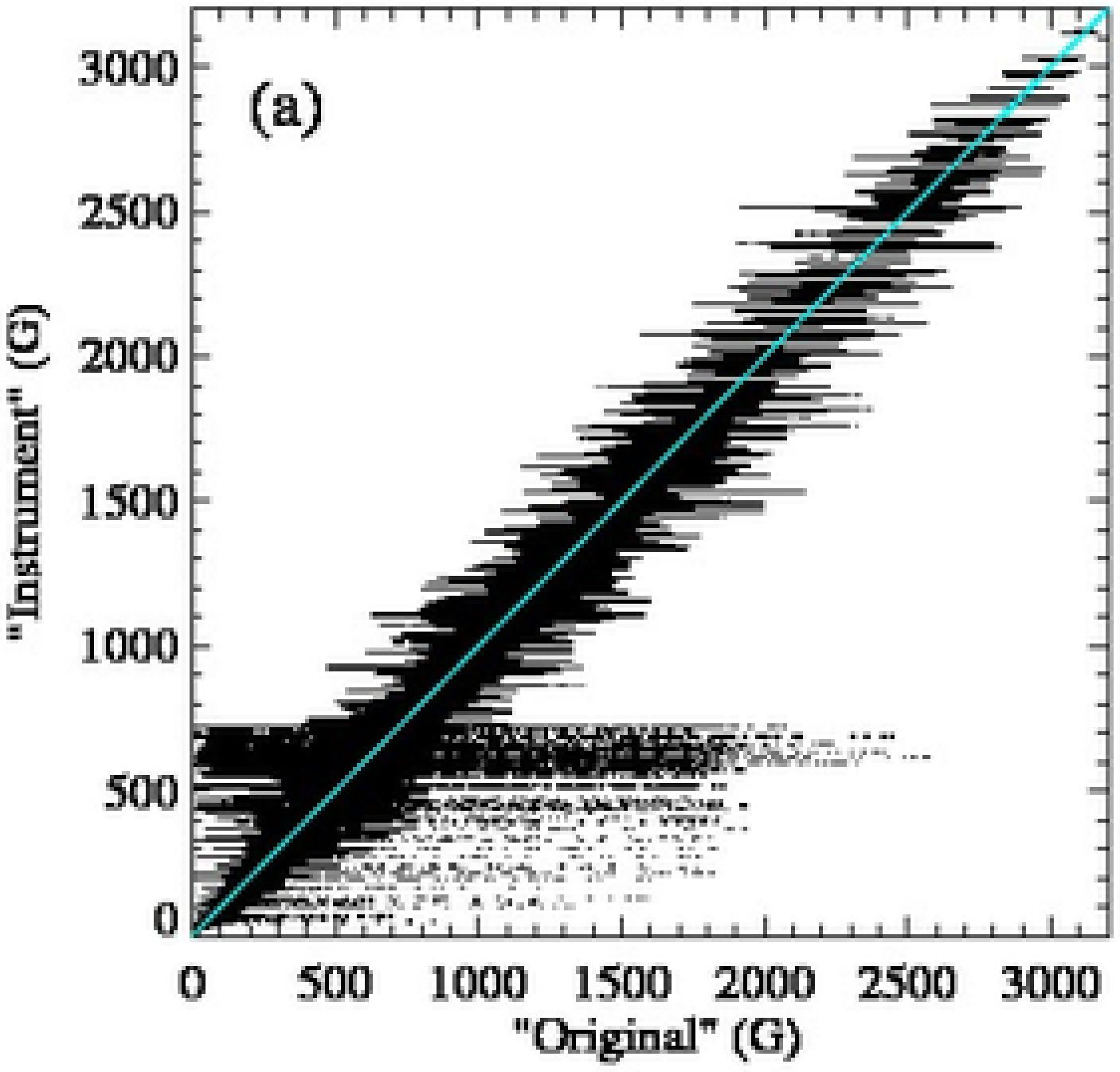}
\hspace{-0.7cm}
\includegraphics[width=0.57\textwidth,clip=]{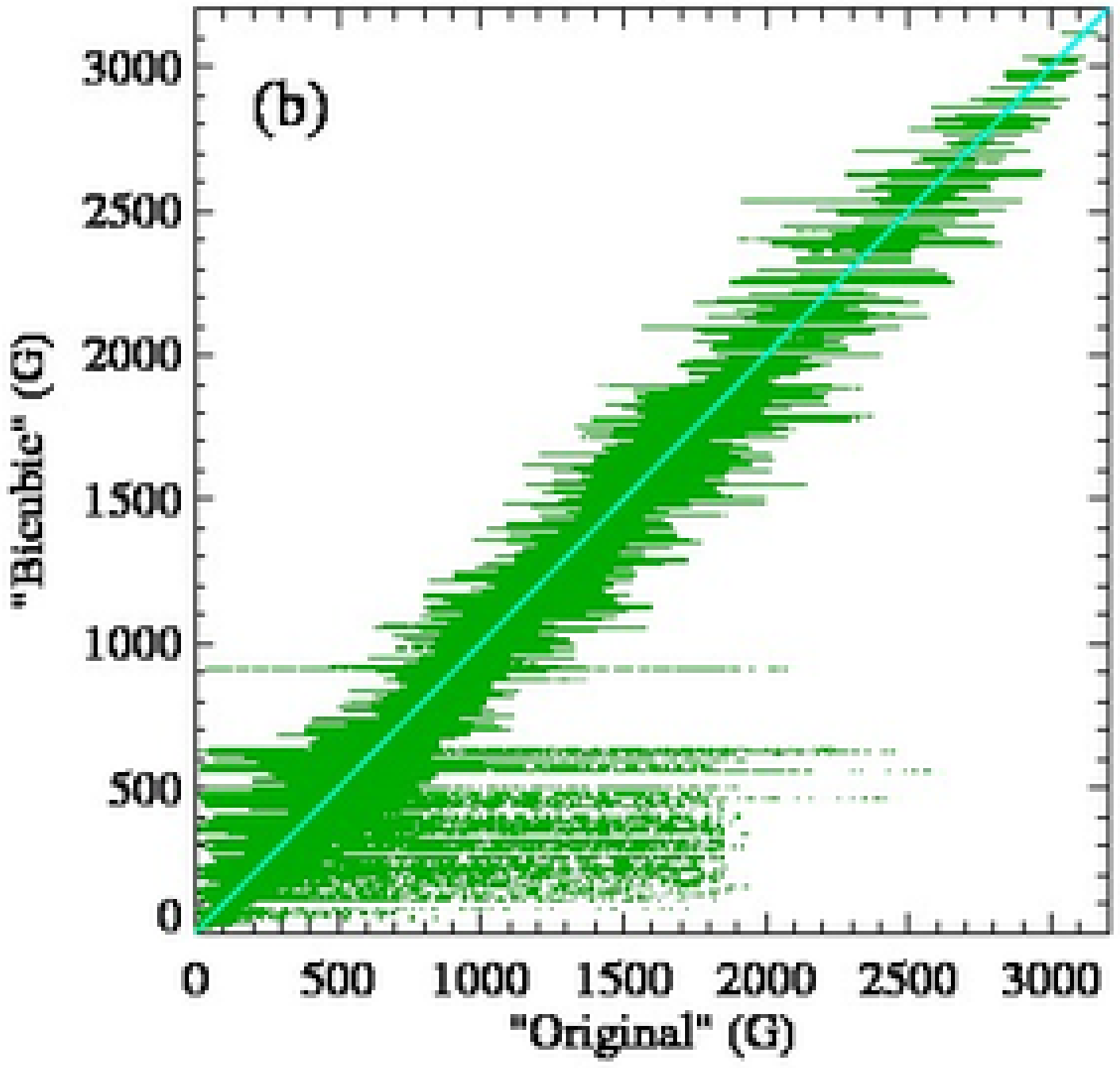}}
\vspace{-0.5cm}
\centerline{
\includegraphics[width=0.57\textwidth,clip=]{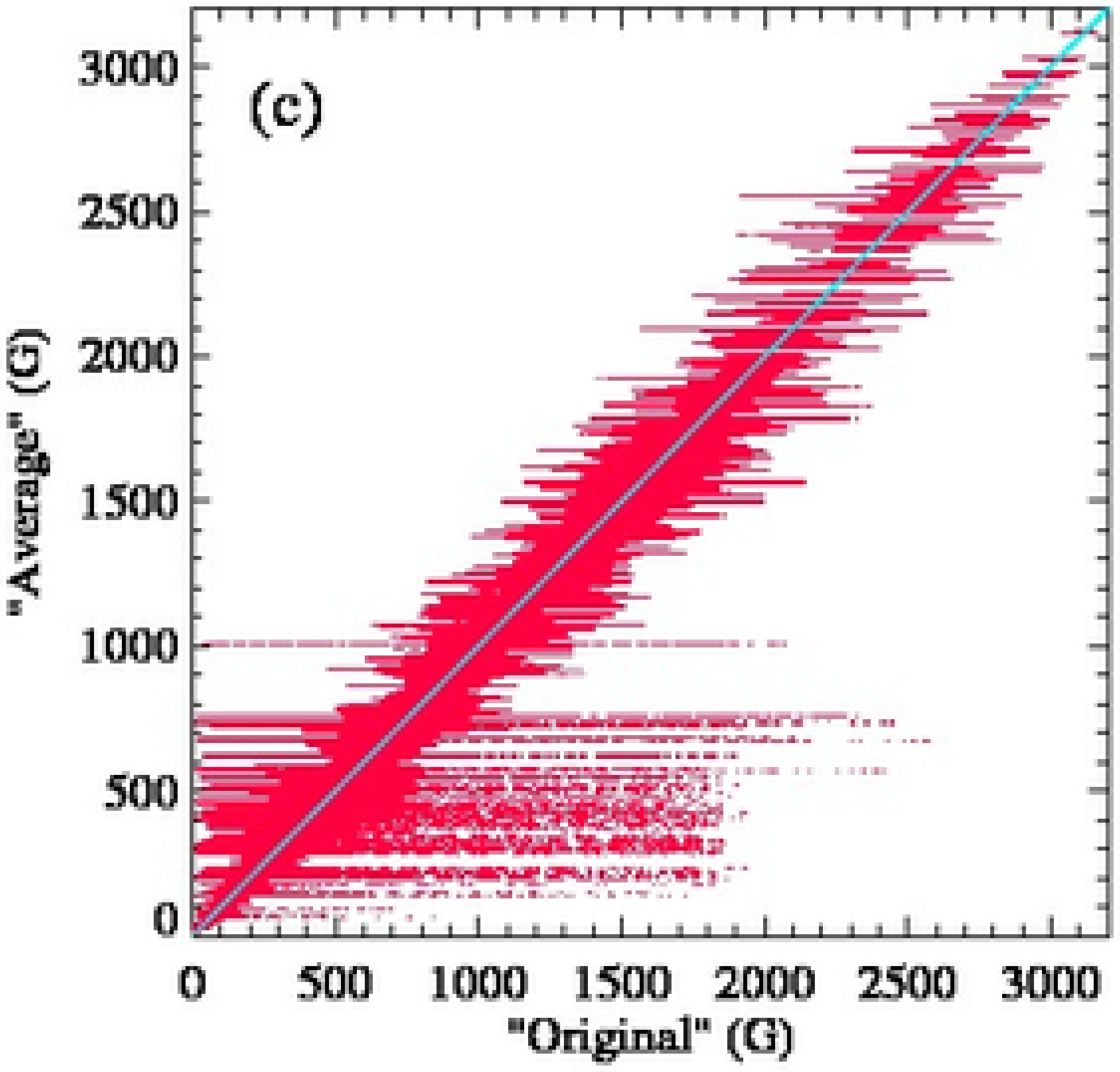}
\hspace{-0.7cm}
\includegraphics[width=0.57\textwidth,clip=]{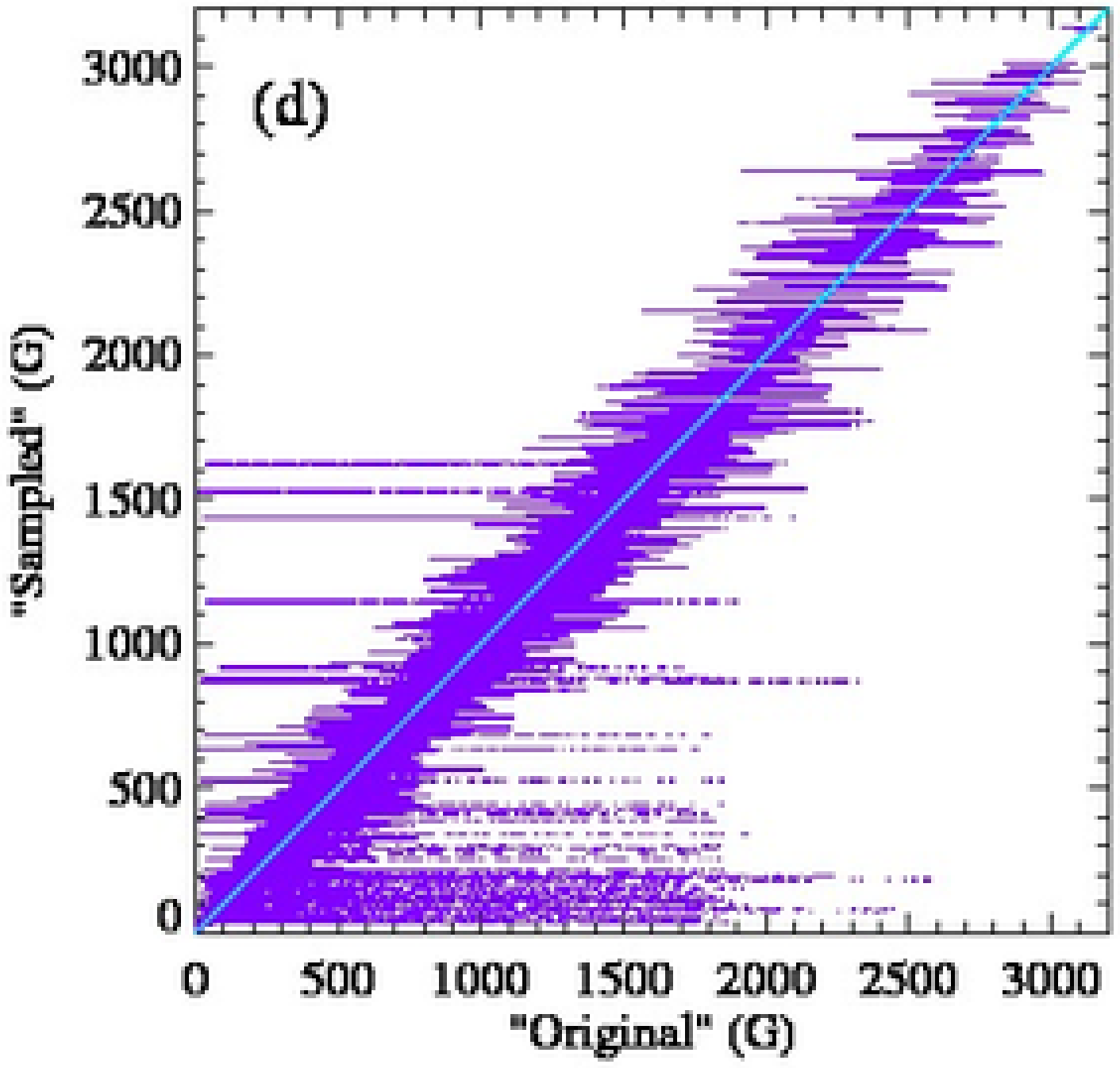}}
\vspace{0.15cm}
\caption{Intrinsic field strength $\B$, comparing the original model magnetogram 
to the bin-by-30
results, for different binning approaches. (a) original {\it vs.}
{\bf ``instrument''}, (b) original {\it vs.} {\color{green} ``bicubic''}, 
(c) original {\it vs.} {\color{red}``average''}, and (d) original {\it vs.}
{\color{violet} ``sampled''}. For all, the $x=y$ line is also plotted
for reference, and on the $x$-axis (``Original'') are plotted
all the values represented by the single resulting
bin-30 pixel in question, whose value is 
plotted on the $y$-axis.  Every other point in the binned magnetogram is 
shown, and every 3rd point of the 900 underlying
values is plotted.
The colors for these plots will be used consistently below.}
\label{fig:flowers_btot_scat} 
\end{figure}

As seen in Figure~\ref{fig:flowers}, spatial rebinning
of any sort produces a boxy, somewhat distorted magnetic field map.
Quantitatively, however, which of the underlying field's properties are
preserved and which are most affected by the change in resolution?

A scatter plot is a good starting place.  In
Figure~\ref{fig:flowers_btot_scat} the intrinsic field strength $\B$ is
compared between the original model and the four ways of binning.  For all
methods, the averaging produces a field that {\it generally} follows the
underlying field distribution; this is reflected in that regression 
analysis returns a near-unity slope ($\gtrsim 0.98$) for each method.
The extremes are lost in what may be termed the ``weak field'' areas (up
to $\approx 1$kG in the binned case) which are in fact highly structured.

Inversions can sometimes fail to return field strength separately from
magnetic fill fraction, especially at low polarization signals.  It has
been shown that the product of these quantities is significantly more
``robust'', meaning easier to retrieve reliably \cite{unnofit} (but see
also \opencite{delToroIniesta_etal_2010}).  Applying the same regression
analysis to the product $f\times\B$ indicates that this is not a cure
for degraded spatial resolution: slopes and standard deviations which
result differ almost imperceptibly, as do the underlying scatter plots,
so we do not show them here.

\begin{figure}[t]
\centerline{
\includegraphics[width=1.0\textwidth]{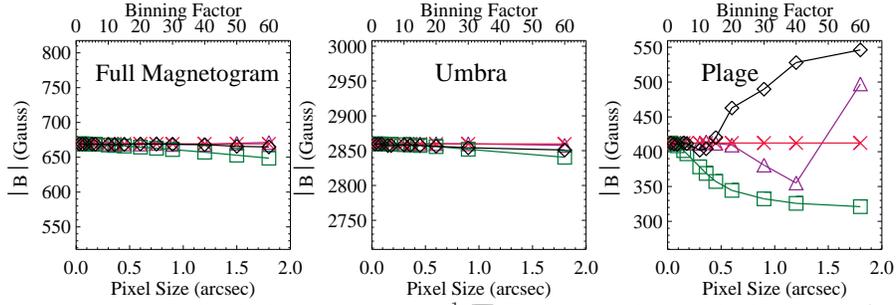}}
\vspace{-0.15cm}
\caption{\small Average intrinsic field strength
$\frac{1}{N} \sum{\B}$
as a function of binning factor (top x-axis), for the four binning methods: 
{\bf ``instrument'' ($\Diamond$)}, {\color{green}``bicubic'' ($\Box$)}, 
{\color{red} ``average'' ($\times$)}, and {\color{violet} ``sampled'' ($\triangle$}).
The three panels show, respectively, the full magnetogram, an ``umbral''
area and a ``plage'' area (see Figure~\ref{fig:flowers}).  
For each binning, $N$ varies but the same sub-area
of the ``Sun'' is covered; when non-integer pixel numbers result, 
that bin factor is omitted.
The original model field is sampled at an arbitrarily-set $0.03\arcsec$,
the resulting ``pixel sizes'' are indicated (bottom x-axis).
For these and all similar plots (except where noted), the y-axis ranges are kept consistent
between the target areas
for direct comparisons.
Here, the effects are minimal for the full magnetogram and the ``umbra'',
but have a much larger magnitude and differ between the binning methods in the ``plage'' area.}
\label{fig:flowers_BB}
\end{figure}

\begin{figure}
\centerline{\includegraphics[width=1.0\textwidth]{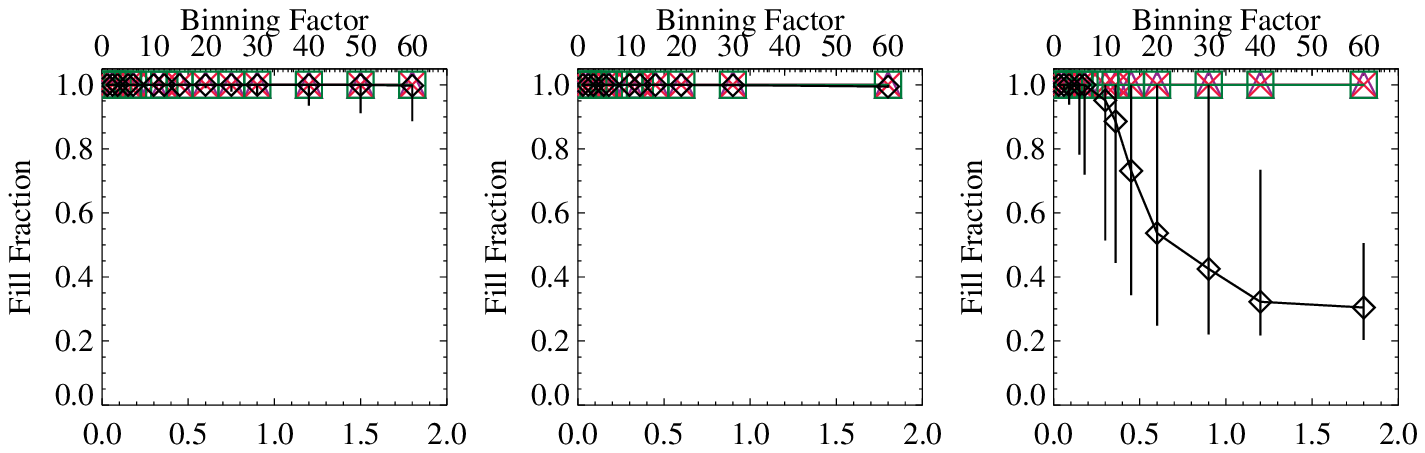}}
\vspace{-0.6cm}
\centerline{\includegraphics[width=1.0\textwidth]{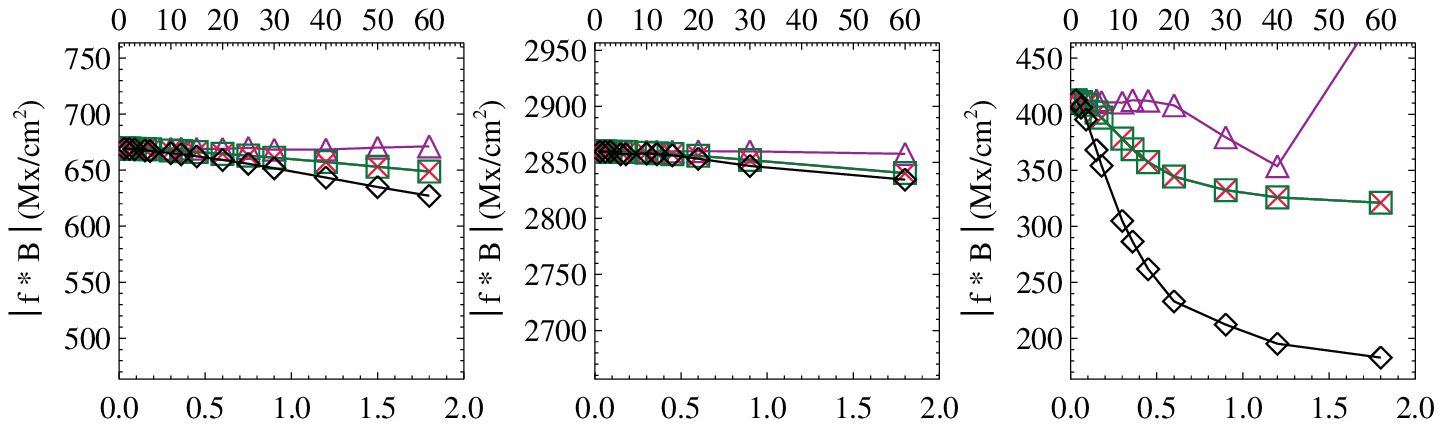}}
\vspace{-0.6cm}
\centerline{\includegraphics[width=1.0\textwidth]{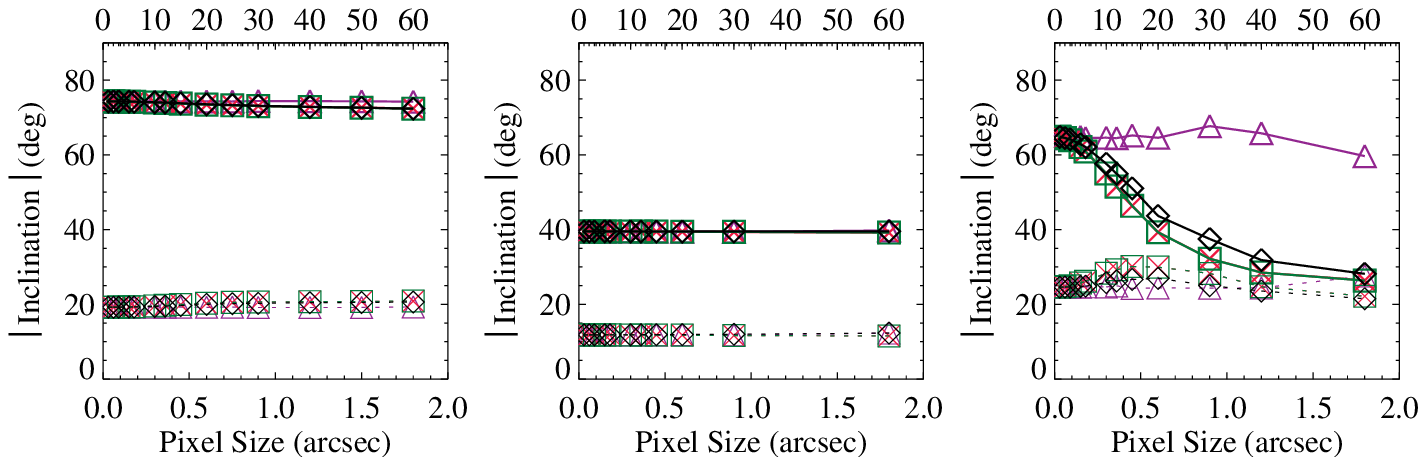}}
\vspace{-0.1cm}
\caption{\small Following the format of
Figure~\ref{fig:flowers_BB}, Top row: median (symbols) and 10th,90th percentiles (displayed
as ``error bars'') of inferred magnetic fill fraction as a function of bin factor.
The three ``post-facto'' approaches consistently return
unity since the original model (and bin-1 inversion)
have unity fill fraction throughout.
Middle row: the average product of the fill fraction and field
strength, $\frac{1}{N} \sum{f\B}$
as a function of binning factor.  
Bottom row: variation of the average inclination angle
with binning factor (thick line-connected curves), $0^\circ$ indicates
(unsigned) fields directed along the line of sight, or pure $\Bl$, and $90^\circ$ indicates field 
perpendicular to the line of sight or pure $\Bt$ (here, $\gamma=\tan^{-1}(\Bt,|\Bl|)$).  Dot-connected
curves indicate the standard deviation of the angle distribution. }
\label{fig:flowers_fffBgamma}
\end{figure}

\begin{figure}
\centerline{\includegraphics[width=1.0\textwidth]{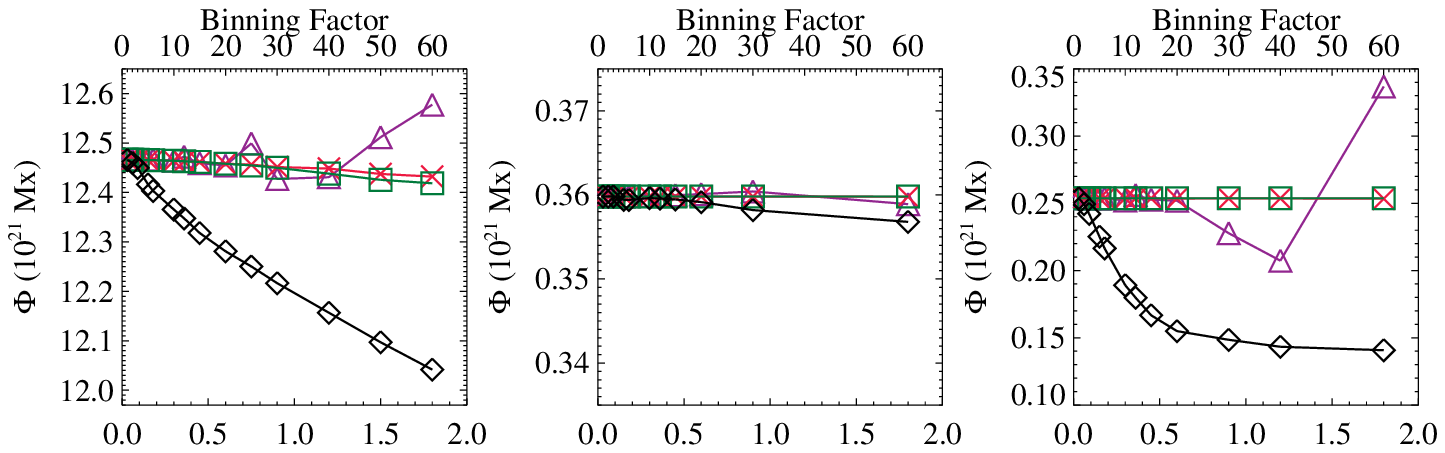}}
\vspace{-0.6cm}
\centerline{\includegraphics[width=1.0\textwidth]{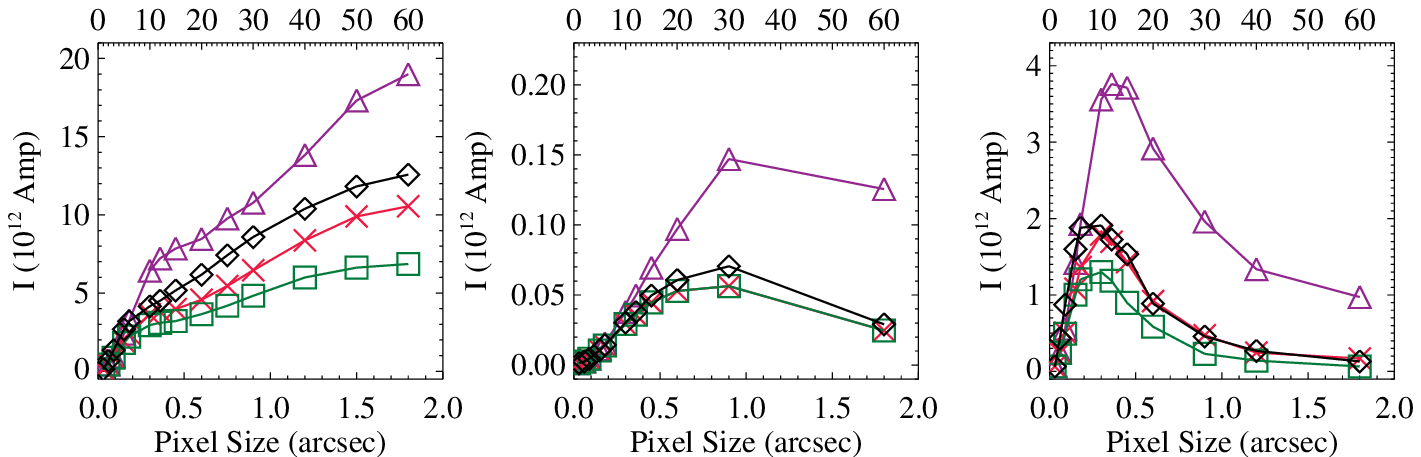}}
\vspace{-0.15cm}
\caption{\small Variation with spatial resolution of parameters often used for characterizing
active regions.  Top: the total unsigned magnetic flux $\Phi=\sum{f|\Bz|\, dA}$.
Bottom: the total unsigned electric current $I = \sum{|\Jz|\, dA}$.  For these plots,
the $y$-axis ranges vary.}
\label{fig:flowers_flux}
\end{figure}

We now examine the inferred magnetic components for the four binning
methods (``instrument'', ``simple'', ``bicubic'', and ``sampled'')
for three target areas (``umbral'', ``plage'', and the full field
of view, see Figure~\ref{fig:flowers}), as a function of different
binning levels.  The nature of this comparison is shown in detail in
Figure~\ref{fig:flowers_BB}.  The intrinsic field strength averaged
over the (sub)-region in question, $\frac{1}{N} \sum{\B}$ is shown as
a function of binning factor for the three target areas.  The results
for the binning methods are shown for each sub-area.  
Comparisons following this format are presented for the magnetic
fill fraction, the product of the fill fraction and field strength,
and the inclination angle distribution (Figure~\ref{fig:flowers_fffBgamma}).
The total unsigned magnetic flux $\Phi = \sum{f| \Bz|\, dA}$
(Figure~\ref{fig:flowers_flux}) is presented, acknowledging the 
somewhat arbitrary assignment of pixel size.
The inferred vertical electric current density $\Jz = C \nabla \times f \Bh$
was computed for the maps using a finite-difference method that employs a 4-point
stencil \cite{precip1} and $C$ includes all the appropriate physical
constants; from this, the total unsigned vertical current $I = \sum {|\Jz|\, dA}$
is presented (Figure~\ref{fig:flowers_flux}) with the same acknowledgment
regarding the assigned pixel size as above.  

The most significant difference between the plage and umbral areas
in the synthetic data is the fact that the former comprises small
scale structure.  The umbral area has essentially one magnetic center,
whereas the plage area contains a few hundred centers that are highly
localized with almost field-free regions separating each center.
The different underlying structure of the field leads to different behavior
at different ``spatial resolutions'', according to the approach.

For field strength (Figure~\ref{fig:flowers_BB}), none of the methods
show dramatic differences in the umbral area; the same is true for the
``full magnetogram''.  In the plage area, the methods behave quite
differently.  Simple rebinning shows absolutely no change with bin-factor, consistent
with its approach of numerically averaging the positive-definite input.
The bicubic approach shows a decrease in average field strength, as
interpolation increasingly underestimates the strong field strength
in the scattered magnetic centers.  The sampling follows the simple averaging
until approximately bin-20 when it decreases, before abruptly increasing at
bin-60.  When the bin-factor is small, the sum over the subset of sampled
points gives a reasonable approximation to the sum over all the (bin-1) points.
As bin-factor increases, the number of sampled points used to represent the sum
decreases, and the result is likely to be increasingly large changes, but with
no consistent trend toward increasing or decreasing with bin-factor.  The
instrument binning in the plage area similarly shows minimal effect until
approximately bin-10, beyond which the average field strength in the plage area
increases.  The polarization-free ``gaps'' between centers begin to be
``contaminated'' with polarization at higher bin factors, and the resulting
average field strength increases, in part because this synthetic plage area is
unipolar. 

The inversion method separately fits for the field strength
and the magnetic fill fraction (Table~\ref{table:terminology}).
The synthetic model is fully resolved, so that for bin-1 all pixels
return unity fill fraction, and hence all ``post-facto'' approaches maintain
unity fill fraction for all bin-factors.  When an inversion is performed on
spatially-averaged spectra, there is almost no effect in the umbral area
(Figure~\ref{fig:flowers_fffBgamma}, top):  the median fill fraction remains
unity.  The situation is very
different in the plage area: the non-unity median and wide {\it range} of
fill fraction returned clearly indicate that worsening resolution leads to
unresolved structures.  The full field of view results reflect a mix of influences from
the ``resolved'' and ``unresolved'' areas in the field of view.

Whether the underlying structures are resolved or not as indicated
by non-unity fill fraction, clearly appears to factor into how 
worsening spatial resolution will affect the field distribution.
The product of fill fraction and field strength (Figure~\ref{fig:flowers_fffBgamma})
which is arguably a better measure of inversion output, 
is the same as the field strength for the ``post-facto'' approaches, but 
shows a dramatic drop under ``instrument'' binning.  The increase in field
strength is more than compensated by a decreasing fill factor, likely as a
result of the intensity weighting of the average Stokes spectra. 

Other effects of note: the distribution of inclination angle
(Figure~\ref{fig:flowers_fffBgamma}, bottom) with worsening spatial
resolution is impacted so as to imply an average orientation closer to
the line of sight in the plage than is originally present, for all but
the ``sample'' approach.  In other words, with worse spatial resolution
the $\Bl$ begins to dominate over $\Bt$, which might be expected given
the lower fractional polarization signal for linear as compared to the
circular polarization.

The total magnetic flux (Figure~\ref{fig:flowers_flux}) is almost
insensitive to bin factor if one uses a post-facto approach, yet plummets
with the instrument approach.  The sampling approach is slightly 
variable, again since the value selected will almost randomly hit
strong or weak signal as bin-factor increases.  Still, the difference is clear:
post-facto binning of any kind does not reproduce the effect of spatial resolution.

The total electric current (Figure~\ref{fig:flowers_flux}) {\it increases} with bin factor overall, with 
a more pronounced effect in unresolved areas than in the unity-fill-fraction
umbral region. 
Recalling that the underlying magnetic model is potential, this somewhat
surprising initial increase and the subsequent decrease in plage areas is due
to an interplay between the less-smooth map (see Figure~\ref{fig:flowers}),
and the finite differences used to calculate the vertical current
(see the discussion in \opencite{ambigworkshop2}); also at play are the influence
of the spatial resolution on the relative strength of the horizontal
component (as seen through the variation in the inclination angle)
and the magnetic fill fraction, which is included when calculating the
vertical current density.  The bicubic approach, which attempts to include
the field structure in the approach, is least affected while the sampling
produces the greatest spurious total current.  Comparing the results
for the umbra and plage sub-area to the full field of view, it is clear that
most of the resulting current arises from unresolved areas such as the ``penumbra-like''
regions that dominate the synthetic model.

\subsubsection{Statistical Tests of Similarity}

The question remains how best to characterize the differences in the
results at different spatial resolutions.  We see from the previous
analysis that the resulting magnetograms do differ, but can they 
adequately describe the underlying field?

\begin{figure}[t]
\centerline{\includegraphics[width=1.0\textwidth]{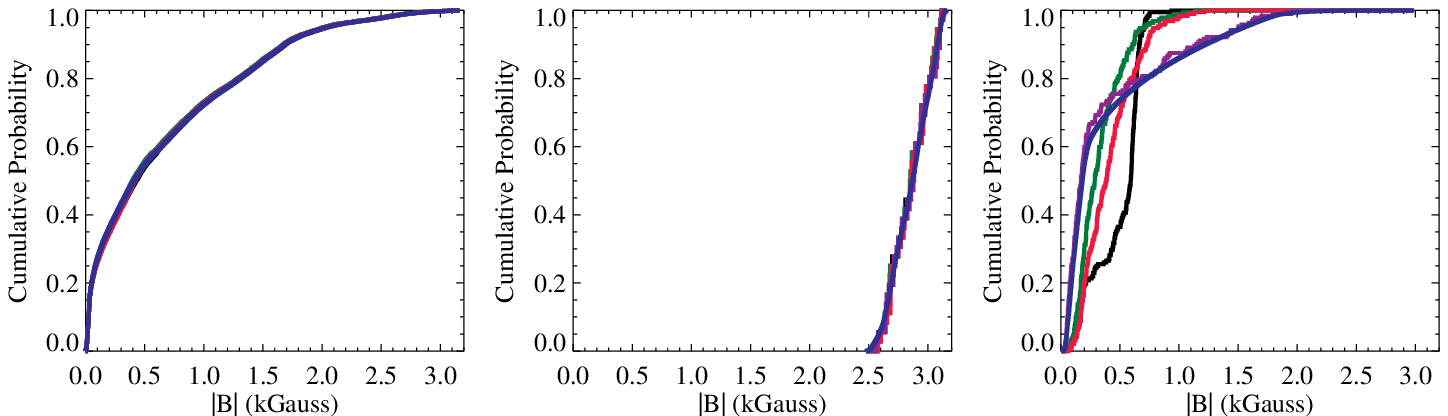}}
\vspace{-0.5cm}
\centerline{\includegraphics[width=1.0\textwidth]{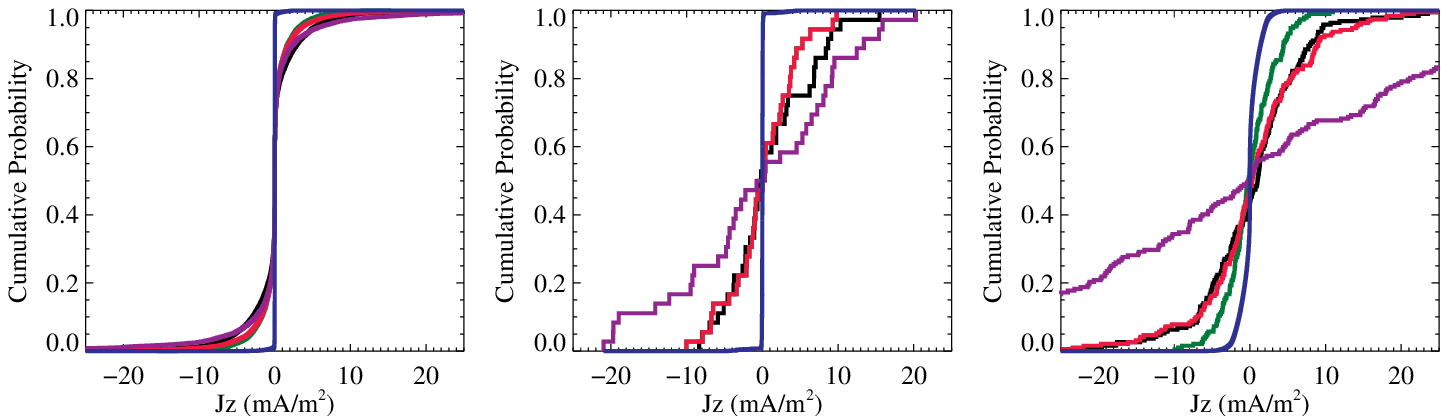}}
\caption{\small Cumulative probability distributions, comparing
that for the full-resolution synthetic map to the bin-30 results, for 
the three fields of view (entirety, ``umbra'', and ``plage'' areas).  
For each, CPD curves are plotted for the {\color{blue} original resolution},
the {\bf instrument} method, and the {\color{green} bicubic}, {\color{red} average}, {\color{violet} sampled}
post-facto approaches.  
Top: For the intrinsic field strength $\B$; 
Bottom: For the vertical electric current density $\Jz$.}
\label{fig:flowers_cpd}
\end{figure}

We perform Kolmogorov-Smirnoff tests on the distribution of the
resulting field parameters to investigate how well a lower-resolution
map characterizes the highest-resolution map.  The K-S test uses the
cumulative probability distribution (CPD) to compare two samples.
Two parameters result: ``$P$'', the probability of rejecting the null hypothesis,
and the ``$D$''-statistic, which measures the maximum difference between
the two CPDs.  In this case the null hypothesis can be stated, ``The two
samples arise from the same population'', the two samples being,
e.g., the map of $\B(x,y)$ from the full-resolution data and the map
from a binned-resolution magnetogram.  It should be remembered that for a given
K-S D-statistic, the KS-probability statistic is extremely sensitive to changes
in the sample sizes, which is very much the case when the bin factor becomes
large.

Comparisons of the CPDs for field-strength and vertical current density
(Figure~\ref{fig:flowers_cpd}) confirm that the widest differences
imposed at the bin-30 level occur in the plage area.  Other parameters
(inclination angle, etc.) show similar behavior.  The umbral area and
the similarity between the CPDs there and the full magnetogram would lead
us to believe (correctly, as demonstrated in Figure~\ref{fig:flowers_fffBgamma})
that for this model, the full magnetogram area is dominated by areas of high
fill-fraction while still containing areas of unresolved highly structured
field.

\begin{figure}
\centerline{\includegraphics[width=1.0\textwidth]{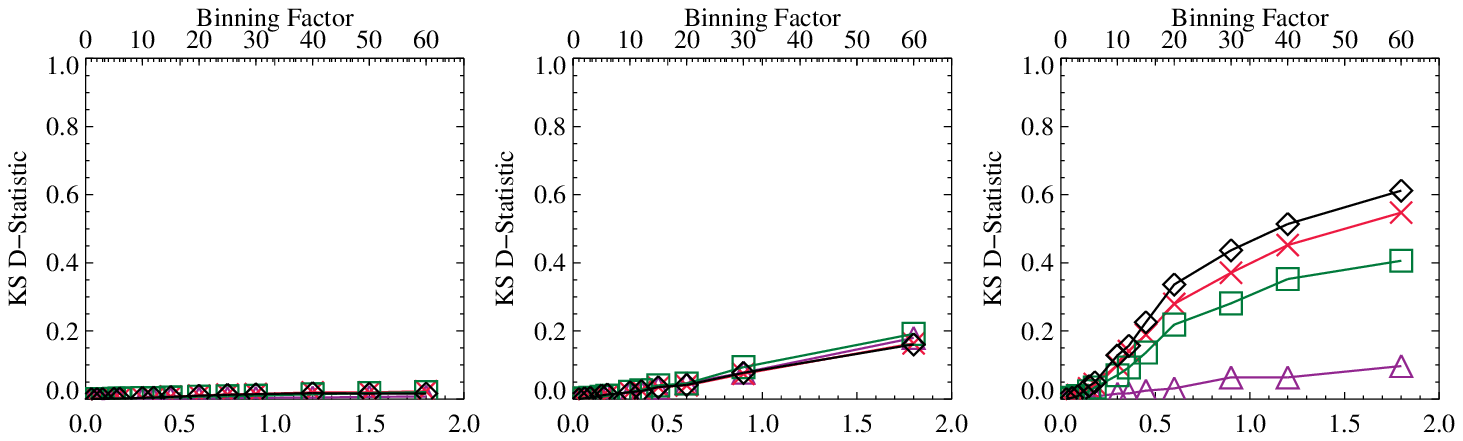}}
\vspace{-0.6cm}
\centerline{\includegraphics[width=1.0\textwidth]{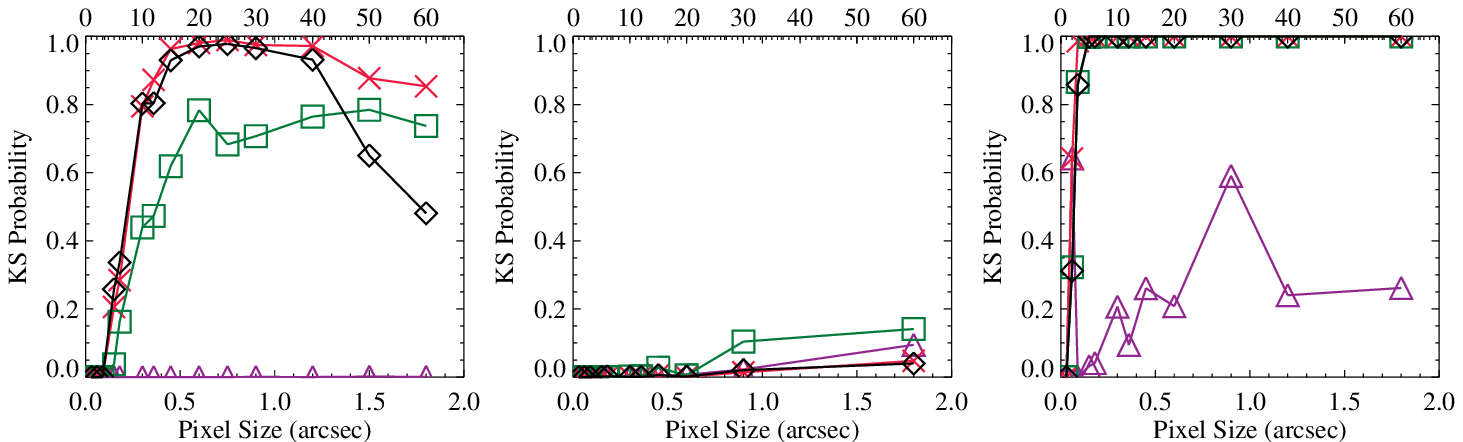}}
\vspace{-0.1cm}
\caption{\small Again for the three fields of view, summaries of the
Kolmogorov-Smirnoff tests as a function of binning factor, for the 
field strength $\B$.  {\bf Top:} the ``D''
statistic and {\bf Bottom:} the probability ``P'' that the
two samples considered are {\it different} (see text).
Shown are four curves, original resolution {\it vs.}
{\bf ``instrument'' ($\Diamond$)}, {\color{green}
``bicubic'' ($\Box$)},
{\color{red} ``average'' ($\times$)}, and 
{\color{violet} ``sampled'' ($\triangle$}) magnetograms.}
\label{fig:flowers_ks_btot}
\end{figure}

For the distribution of field strength, the D-statistic
(Figure~\ref{fig:flowers_ks_btot}) is dominantly zero for the full field of
view, and increases only slightly with worse spatial resolution in the umbra.
However, it is significantly non-zero for the ``plage'' area, reflecting that
all bin factors show the same behavior seen in detail in
Figure~\ref{fig:flowers_cpd}.  The smallest D-statistic in the plage area comes
from the sampling approach; the greatest from the instrumental approach.
The probabilities of rejecting the null hypothesis are mixed but generally
close to unity for the plage area, with systematically lowest probabilities
for sampling, as expected.  

\begin{figure}
\centerline{\includegraphics[width=1.0\textwidth]{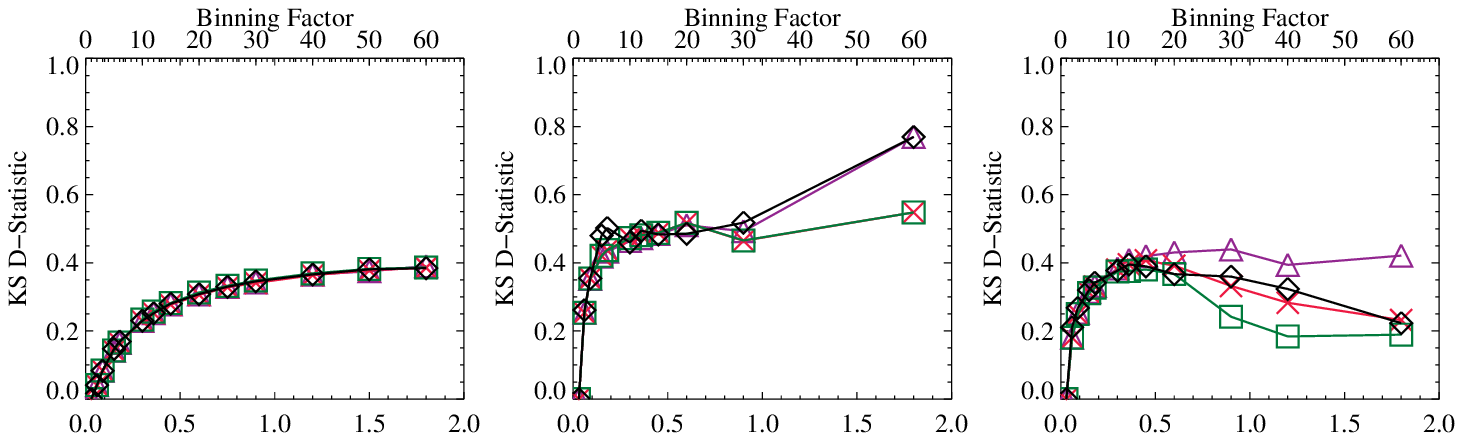}}
\vspace{-0.6cm}
\centerline{\includegraphics[width=1.0\textwidth]{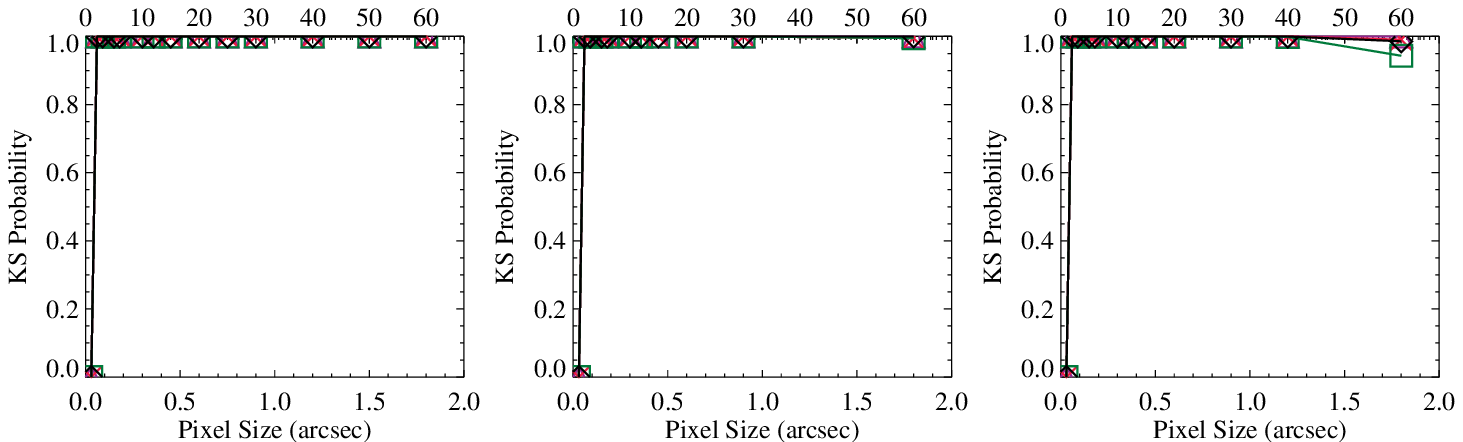}}
\vspace{-0.1cm}

\caption{\small Summaries of the
Kolmogorov-Smirnoff tests as a function of binning factor, for the vertical electric
current density $\Jz$.  Format follows Figure~\ref{fig:flowers_ks_btot}.}
\label{fig:flowers_ks_jz}
\end{figure}

For the distribution of $\Jz$, a quantity derived by taking derivatives of the
field distribution, the K-S D-statistic (Figure~\ref{fig:flowers_ks_jz}) is
significantly nonzero for all three sub-areas and all methods at all spatial
resolutions.  The KS-probability is consistently unity; this bodes ill for the
possibility that unresolved magnetograms recover the underlying distribution of
field or vertical current.

To summarize these results, in areas such as this model ``umbra'', the
underlying field varies little and the inferred fill fraction is consistent 
with it being ``resolved''.  It can be argued that through a wide
range of spatial resolution, the inferred field distribution represents
the underlying field.  The situation for highly structured underlying
field is very different: areas of low- and mixed- fill-fraction imply that
the field is not resolved.  It is fairly clear that instrumental effects
on the spectra result in a substantively different field distribution,
and the implied structures should be treated with much less confidence.
And, with all caveats acknowledged due to the use of synthetic data,
we find that in general, inferring the distribution of the vertical
current is very susceptible to the effects of spatial resolution.

\section{Demonstration: Real Data, Revisited}

One may always argue that synthetic data constructed to demonstrate
a particular effect may not represent observational ``truth''.
Hence, we perform the same exercise using data from
the Solar Optical Telescope SpectroPolarimeter aboard the {\it Hinode} mission
\cite{hinode_sp}.  While the data from this instrument are arguably not
the highest resolution spectropolarimetric data available, the temporal
and spatial consistency coupled with very good resolution in both spatial
and spectral dimensions make these data ideal for this purpose.

We chose the 18:35 UT scan of 30 April 2007 scan of NOAA Active Region (AR) 10953, observed at
S09.5, E11.5 ($\mu = 0.98$), which was a ``normal'' scan that approximately matches
scan-steps to the slit width and does not perform any on-board summation.
The field of
view includes a sunspot and plage area sufficient for this purpose.
The pixels are not exactly square, and are not interpolated to be square, but treated as
unequal in dimension for all analysis;  we do, however, use the average
of $0.15\arcsec$ when referring to general pixel size.

\begin{figure}
\vspace{-0.5cm}
\centerline{
\hspace{-0.50cm}
\includegraphics[width=0.58\textwidth]{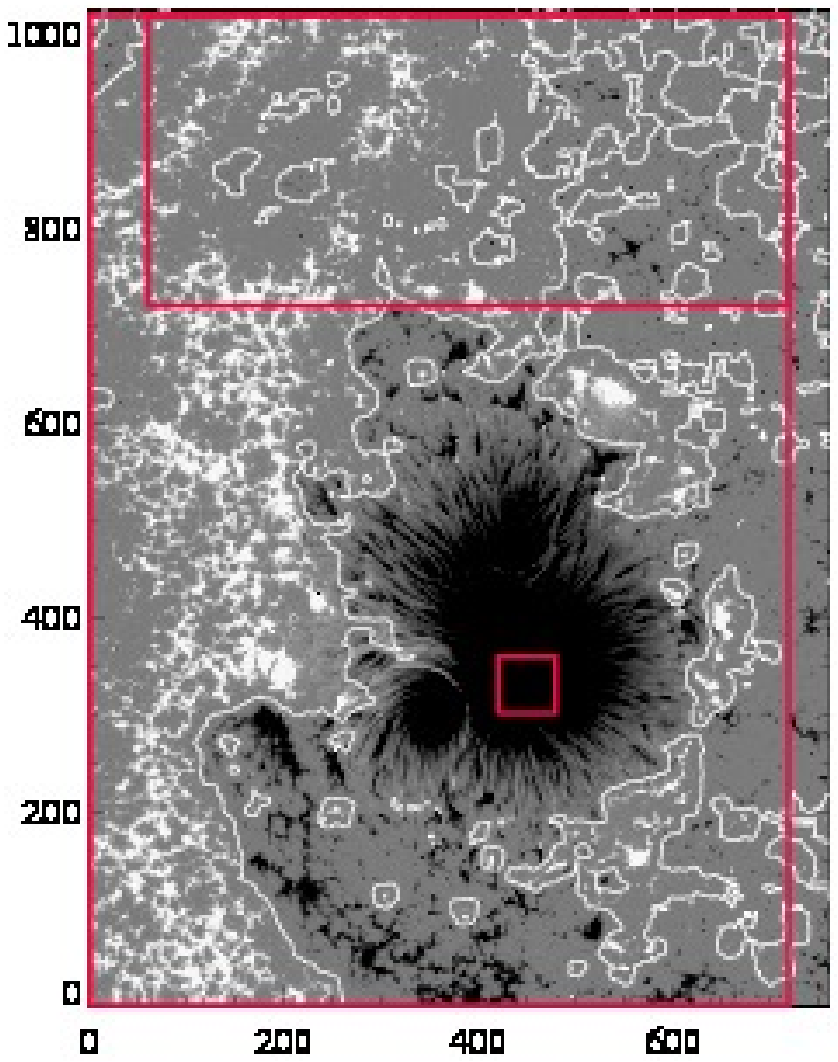}
\hspace{-1.25cm}
\includegraphics[width=0.58\textwidth]{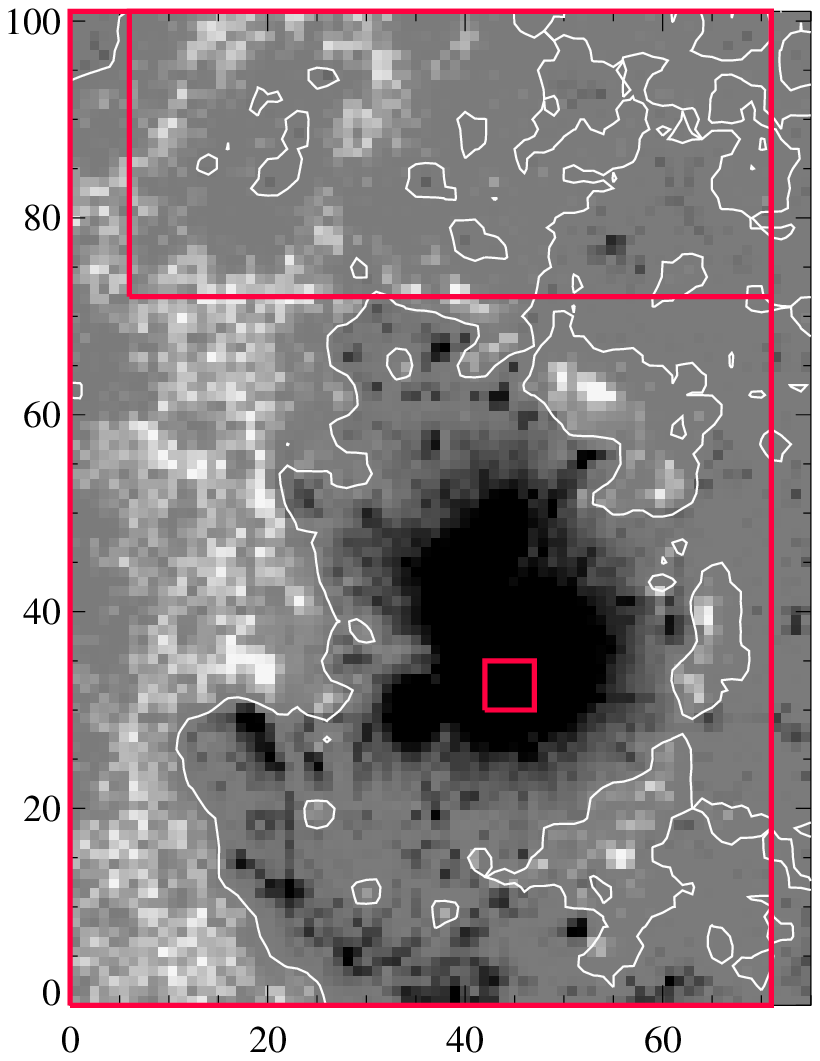}}
\vspace{-0.25cm}
\caption{\small The $\Bl$ component inferred by {\it Hinode}/SP, 
scaled to $\pm 1000 {\rm Mx\,cm}^{-2}$, for NOAA Active Region 10953 observed
18:35UT 30 April 2007.  Left: full-resolution data, with 
original dimensions $762\times1024$ and $0.15\arcsec$ pixels size.  Boxes
indicate the sub-regions highlighted in the later analysis, 
``umbra'' ($60\times60$ pixels), ``plage'' ($660\times300$).  In addition,
as shown the ``full field of view'' is slightly trimmed (to $720\times1020$) to ensure
integer divisibility by a range of factors.   Right: same, after ``instrument''-binning
by a factor of 10, to $1.5\arcsec$.} 
\label{fig:hinode} 
\end{figure}

An approach parallel to that described above was used to treat the
{\it Hinode}/SP data, albeit beginning with the 
fully-calibrated Level-1D $[I,~Q,~U,~V]$ Stokes spectra 
\footnote{\url{ http://sot.lmsal.com/data/sot/level1d/}}.
In this case there is already photon noise present in the data,
and the demodulation is performed on-board.  In the context of the
Poisson-statistics (see Appendix~\ref{sec:photonnoise}), the implications
are that we cannot exactly model the effects of different apertures.
Without the ``raw'' observed modulated $I\pm P,~P \in [Q,~U,~V]$ spectra
and the different contributing realizations of noise, information has
already been lost, and manipulating the demodulated pure $[I,~Q,~U,~V]$
spectra is equivalent to re-constructed mixed-polarization states.
The manipulated (averaged spatially by summation) spectra will present
with lower noise than would actually be the case, but the primary effects
of spatial resolution modeling will still be apparent.

The binned spectra were written in the ``ASP'' format (with a
reformatter courtesy B. Lites, HAO/NCAR), and inverted using the HAO/NCAR
Milne-Eddington inversion code ``sss-inv'' (\opencite{sl87}; \opencite{ls90}; \opencite{litesetal93},
with minor modifications for {\it Hinode}/SP specifics, again courtesy
B. Lites, HAO/NCAR).\footnote{Implementation details: $[I,~Q,~U,~V]$
weighting:$1/[100,1,1,10]$, fill-fraction solved, initial guess
via ``genetic'' algorithm optimization, all pixels inverted (no
minimum-polarization threshold), ``scattered light'' profile determined
where $\sum{|P|} < 0.4\%$} The full-resolution data were subjected
to the same reformatting and inversion (without binning) to ensure a
consistent comparison.  The ``ME0'' minimum-energy code was used in
a consistent manner for ambiguity-resolution for all data, and $\Jz$
was calculated in exactly the same manner as for the synthetic data.
A sample binned magnetogram is shown in Figure~\ref{fig:hinode}.

\begin{figure}[t]
\centerline{
\hspace{-0.25cm}
\includegraphics[width=0.45\textwidth]{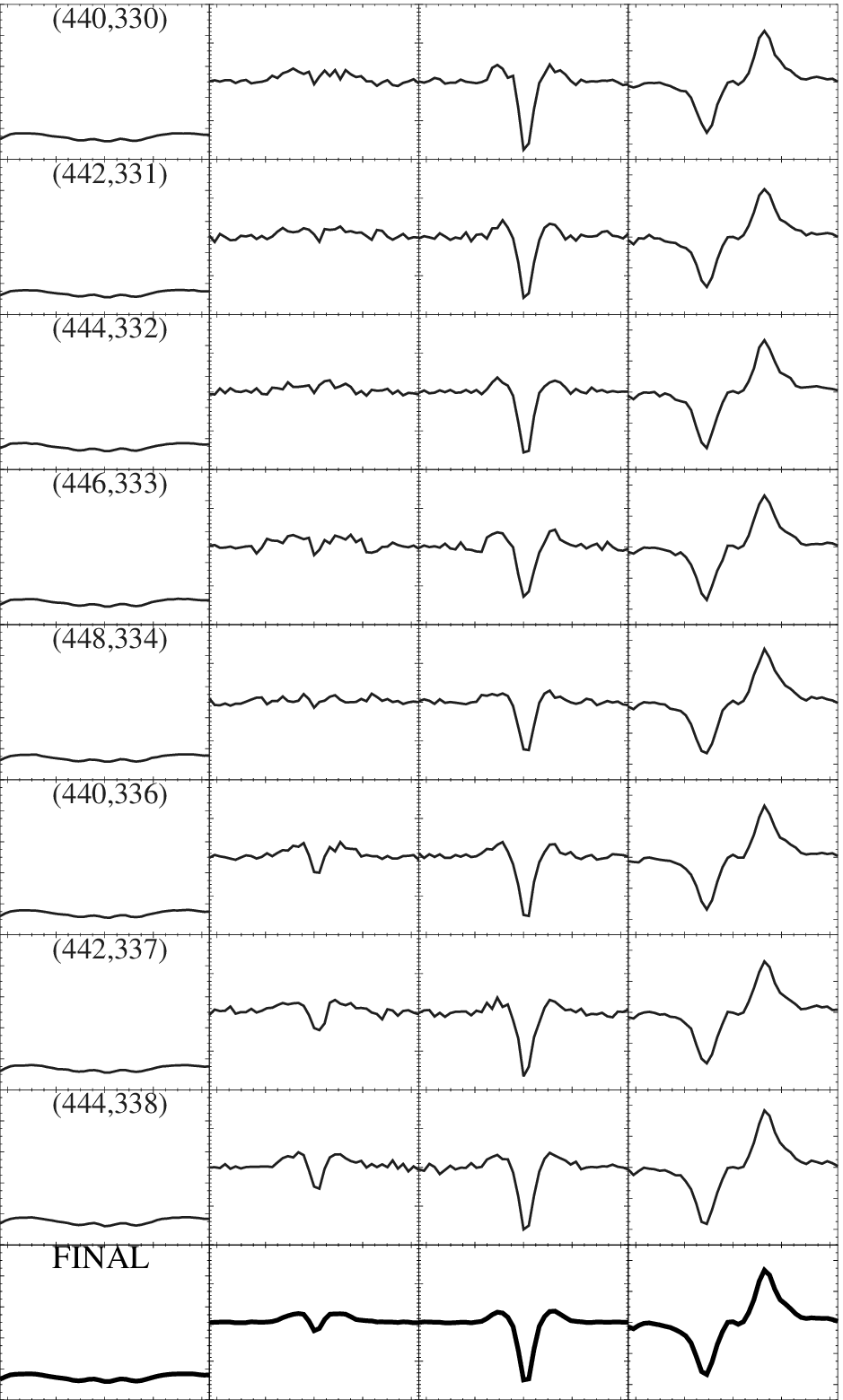}
\hspace{0.25cm}
\includegraphics[width=0.45\textwidth]{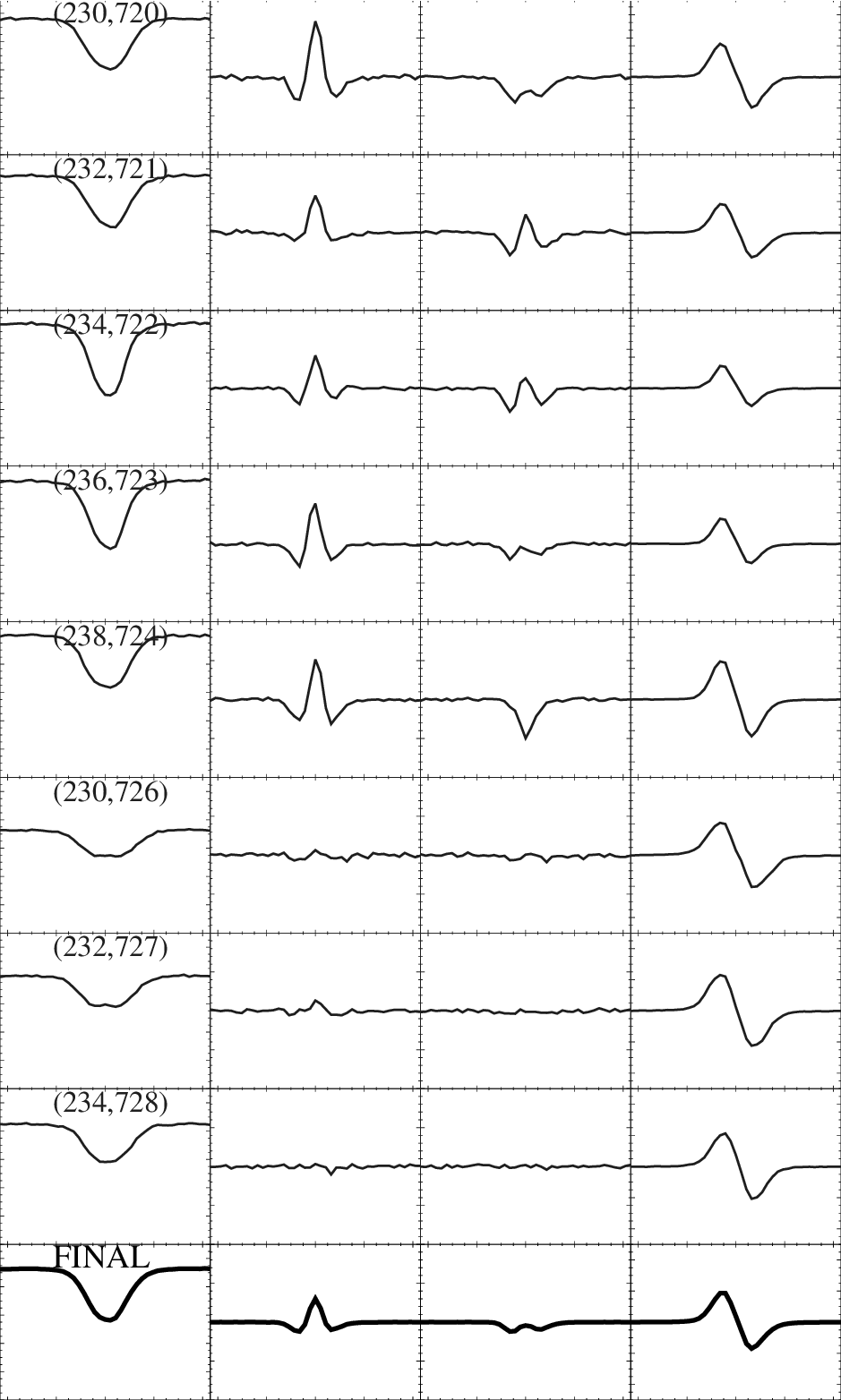}}
\caption{\small  Left column: Eight samples of emergent Stokes
$[I,~Q,~U,~V]$ spectra, from a small patch (10$\times$10 pixels) of the
full-resolution {\it Hinode}/SP map, centered in the sunspot umbra at [445,~335]
in Figure~\ref{fig:hinode}(left).  Stokes $[I,~Q,~U,~V]$ are plotted
left-right with ranges: $I:[0,1],~Q,~U:[-0.2,0.2],~V:[-0.5,0.5]$,
the pixel coordinates (of the original data) are also shown.  
Left, bottom: the resulting ``FINAL'' $[I,~Q,~U,~V]$ after
averaging the 100 underlying emergent polarization spectra, plotted
on the same scale.  Right column: Same as left set, but for a
10$\times$10 pixel area centered in the ``plage'' area, at [235,~725]
in Figure~\ref{fig:hinode}~(left).  For these plage data, the ranges
are $I:[0,1],~Q,\&~U:[-0.1,0.1],~V:[-0.5,0.5]$.}
\label{fig:hinode_spectra}
\end{figure}

As with the synthetic data, three areas are analyzed: the full field
of view, and then separately two areas, one centered on the sunspot
umbra and another on a plage area to the north
of the sunspot (Figure~\ref{fig:hinode}).  The latter area was chosen
to avoid the emerging filament at the south east edge of the sunspot
\cite{okamoto_etal_08}.  The full scan was trimmed slightly and both
sub-areas were chosen to be evenly divisible for a number of binning
factors.

Samples of the effects of ``instrument''-binning on
emergent Stokes spectra from the {\it Hinode}/SP data are shown in
Figure~\ref{fig:hinode_spectra}.  The umbral sample displays very
consistent Stokes spectra, and a final bin-10 result that closely
resembles any single constituent-pixel's set of spectra.  The noise is
nicely reduced in the binned spectra (although somewhat artificially,
as described above and in Appendix~\ref{sec:photonnoise}).  The plage sample demonstrates 
exactly the effect shown
in Figure~\ref{fig:spectra}, that the constituent spectra are
quite variable, and the resulting binned data reflect an average that
does not represent any single underlying pixel.

\begin{figure}
\centerline{
\includegraphics[width=0.57\textwidth,clip=]{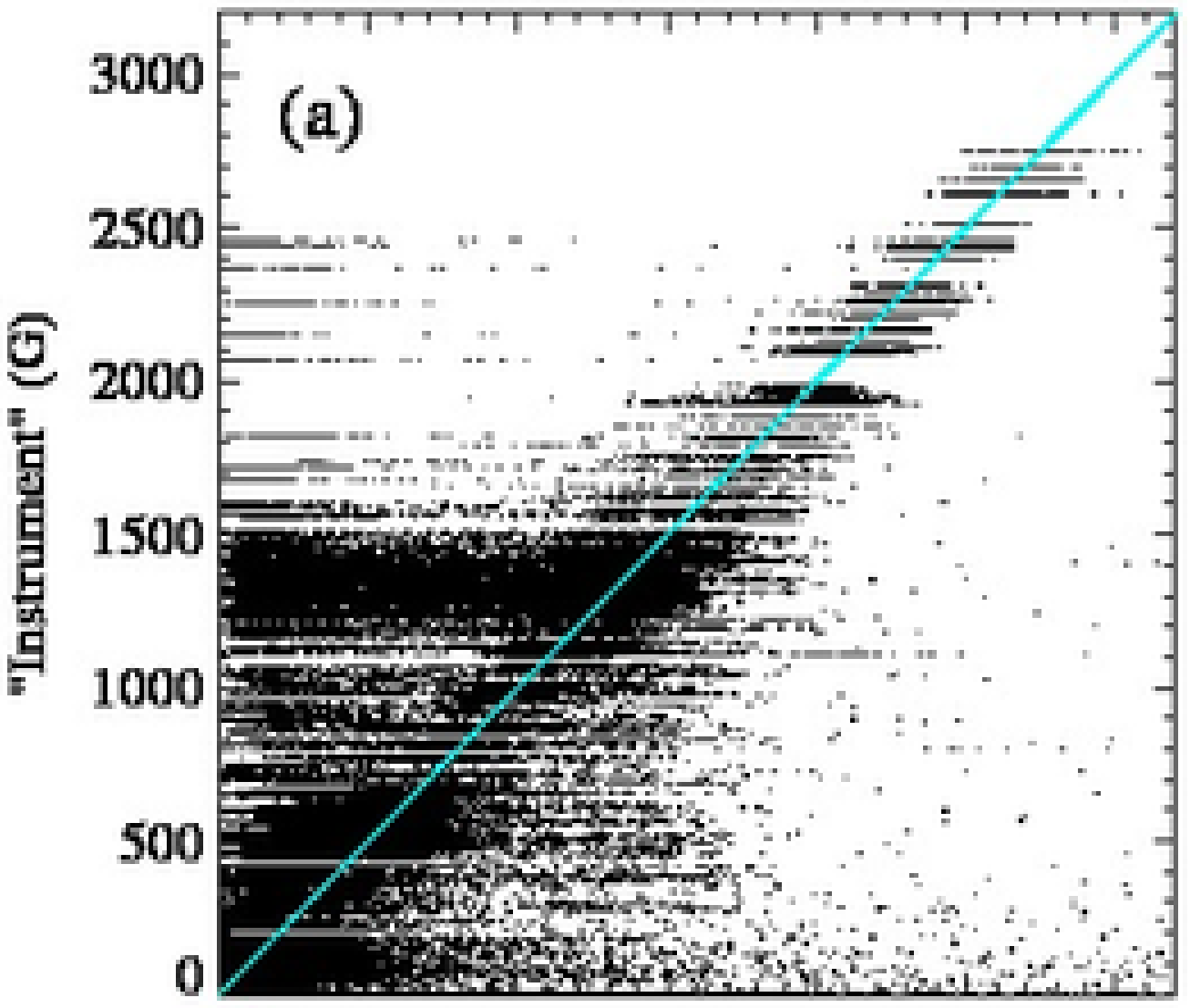}
\hspace{-1.0cm}
\includegraphics[width=0.57\textwidth,clip=]{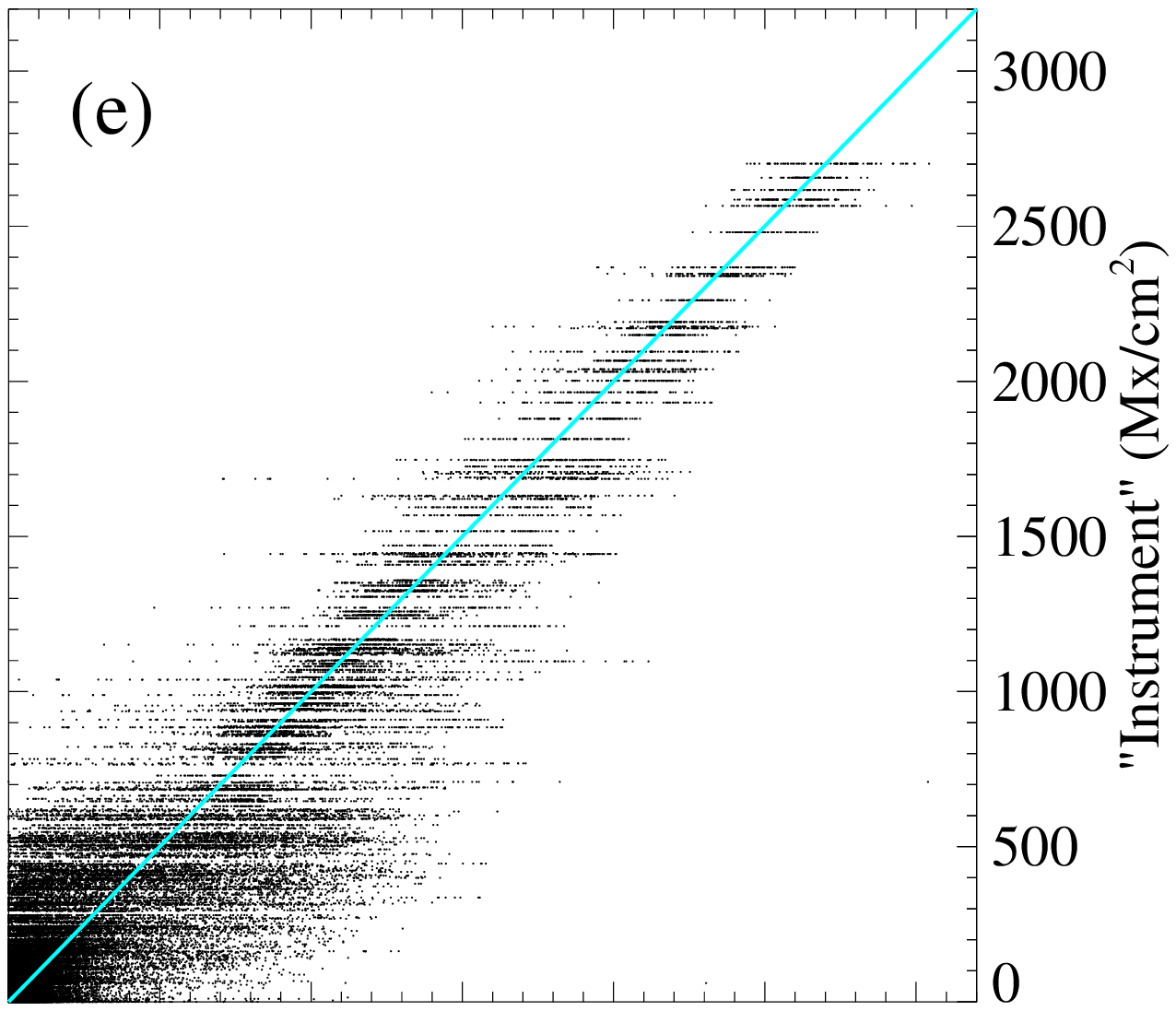}}
\vspace{-0.4cm}
\centerline{
\includegraphics[width=0.57\textwidth,clip=]{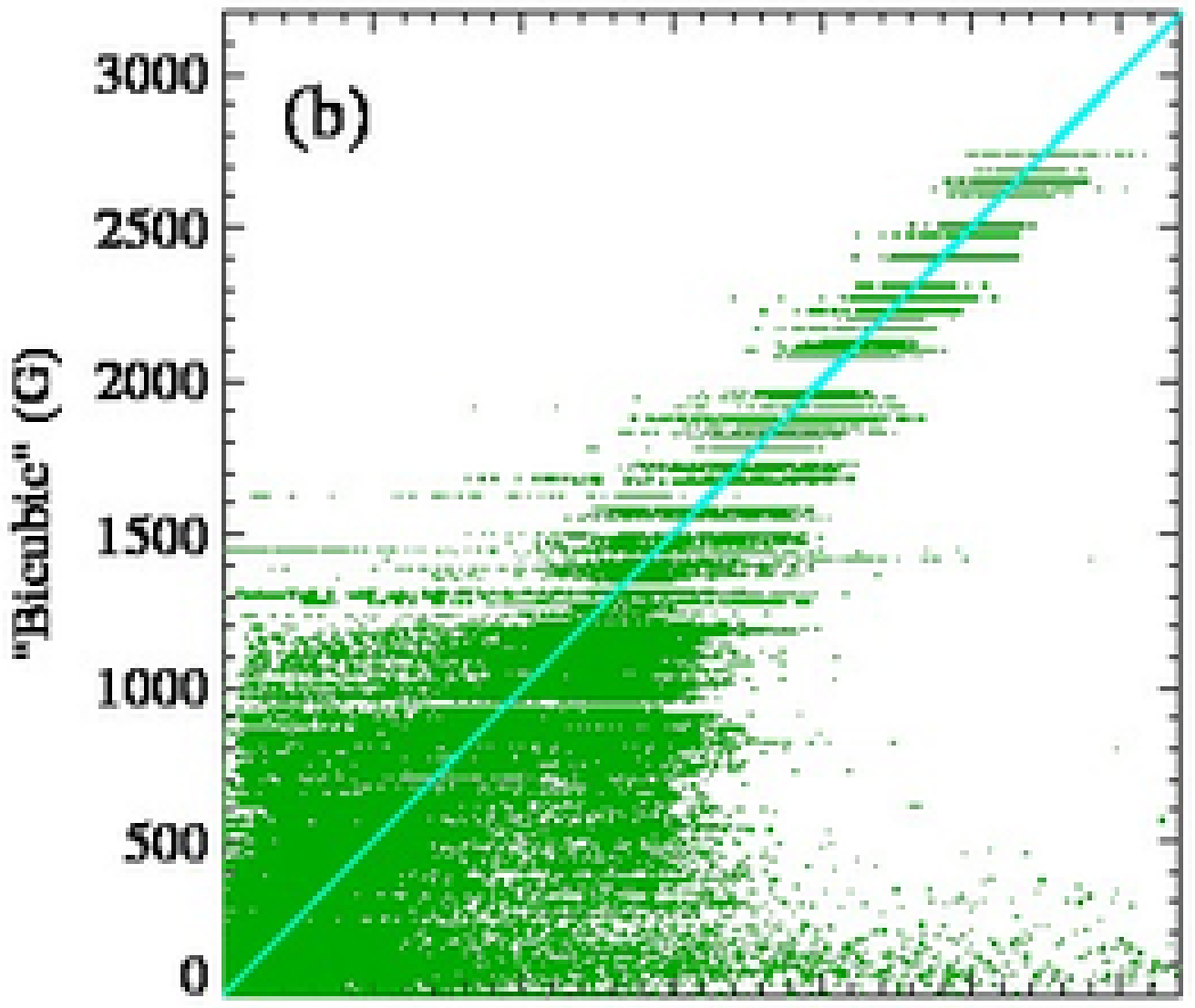}
\hspace{-1.0cm}
\includegraphics[width=0.57\textwidth,clip=]{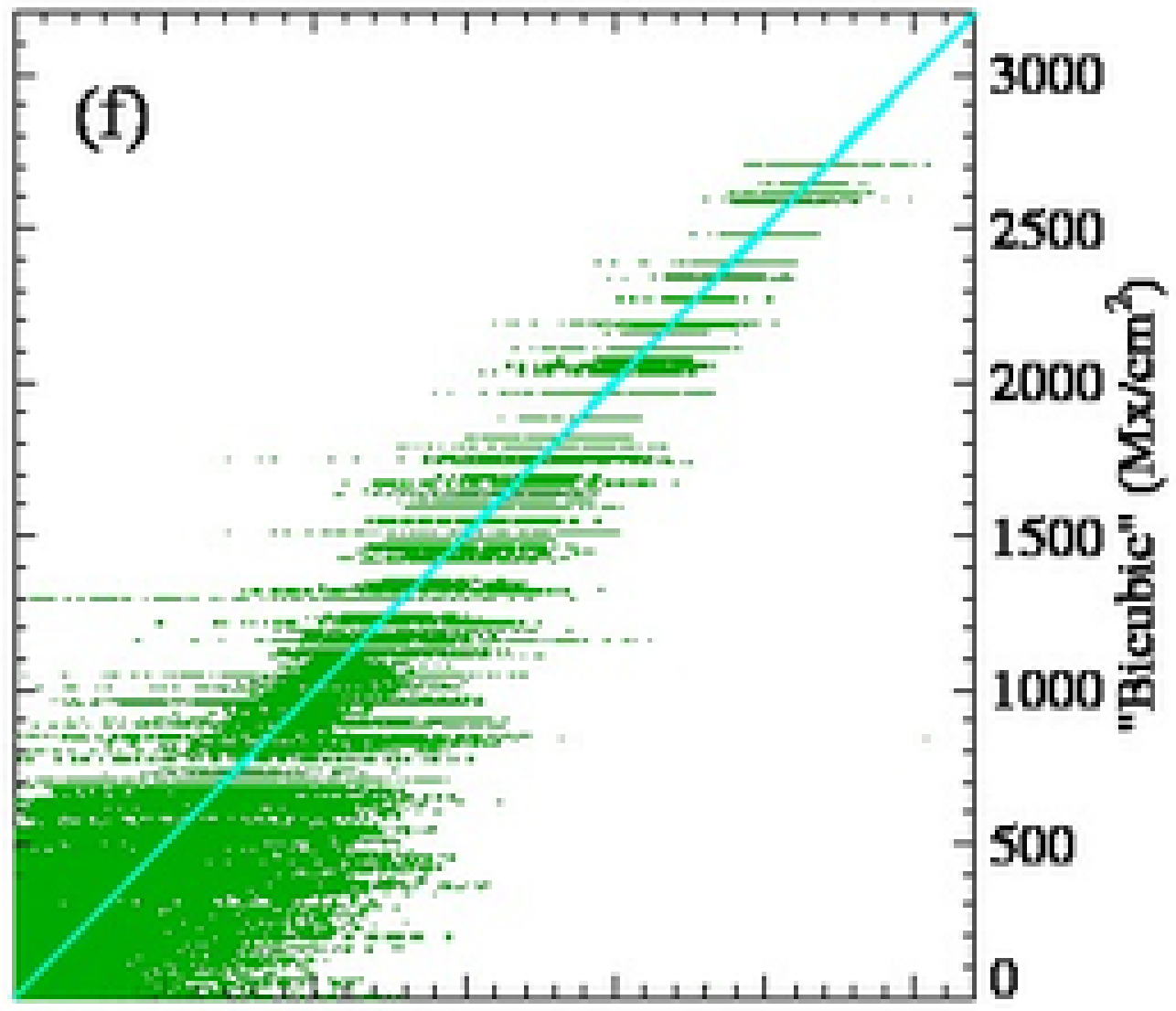}}
\vspace{-0.4cm}
\centerline{
\includegraphics[width=0.57\textwidth,clip=]{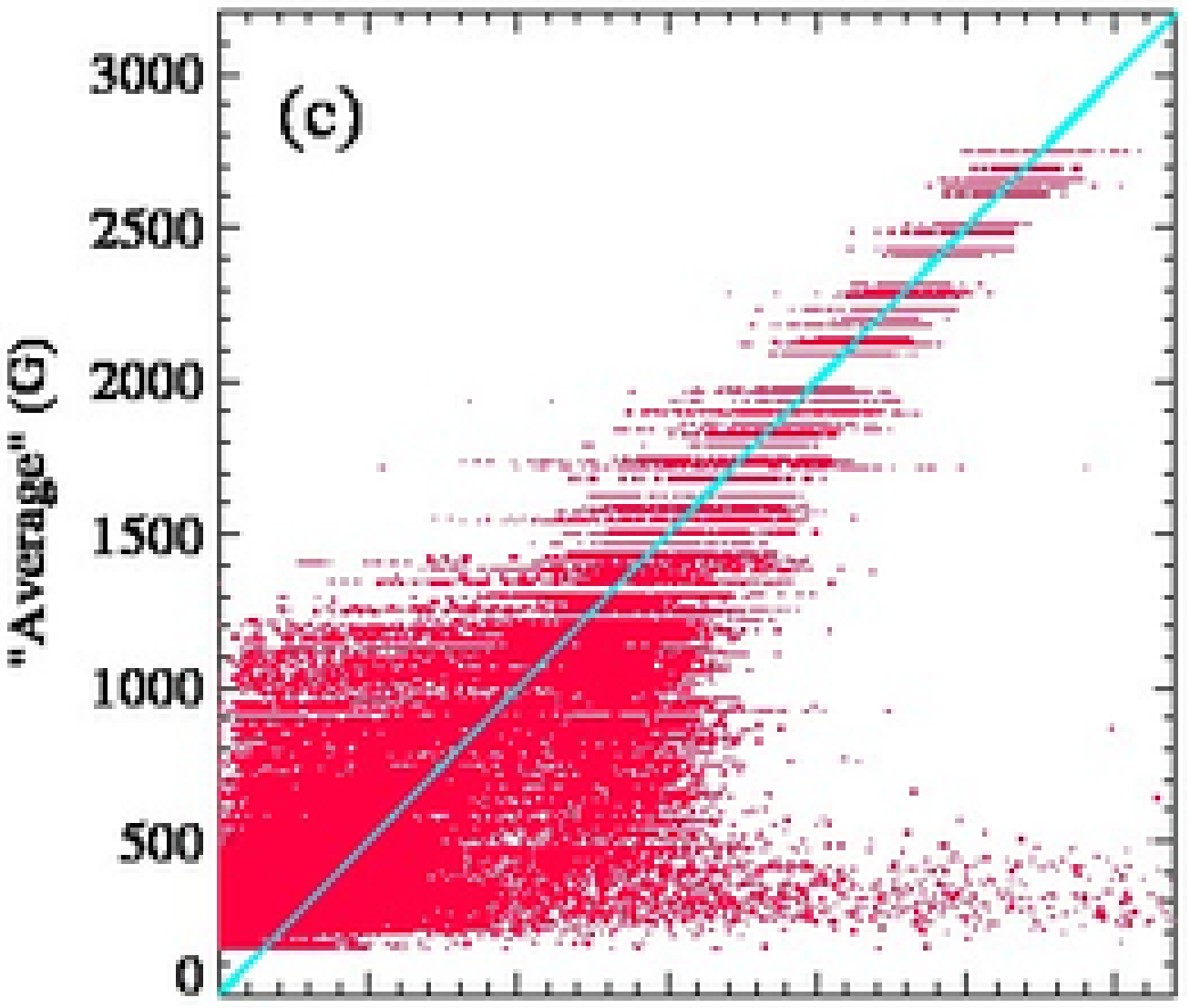}
\hspace{-1.0cm}
\includegraphics[width=0.57\textwidth,clip=]{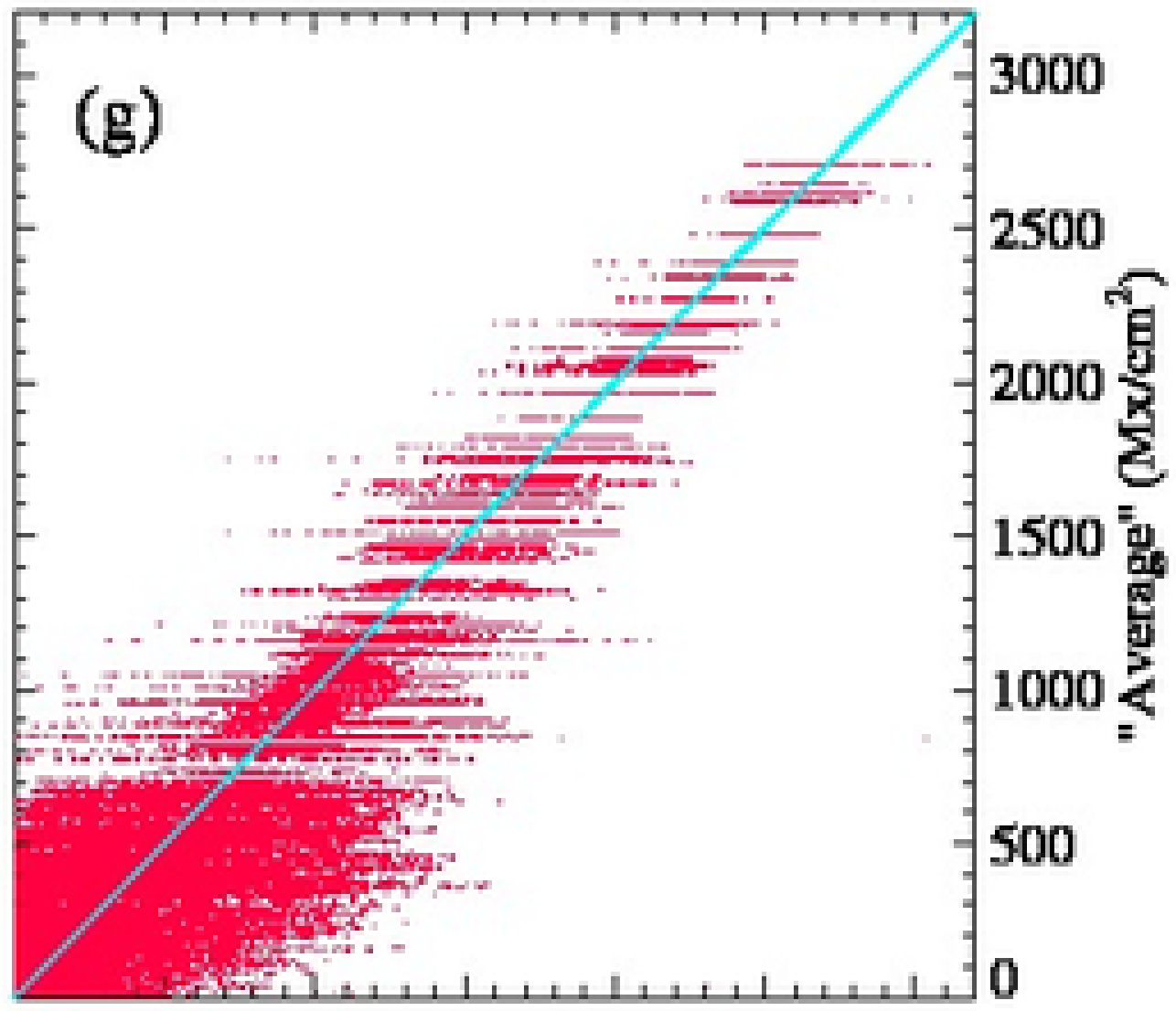}}
\vspace{-0.4cm}
\centerline{
\includegraphics[width=0.57\textwidth,clip=]{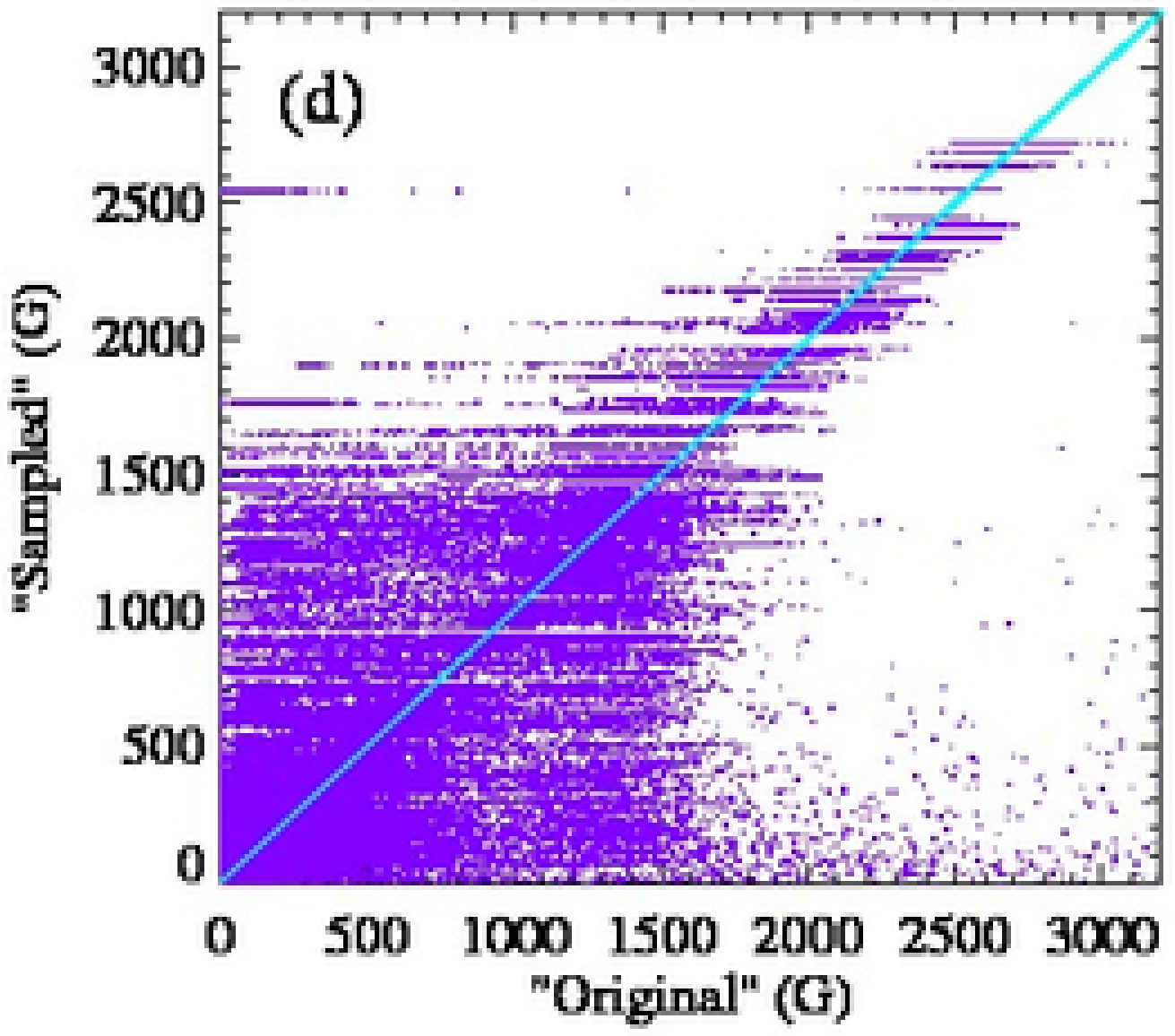}
\hspace{-1.0cm}
\includegraphics[width=0.57\textwidth,clip=]{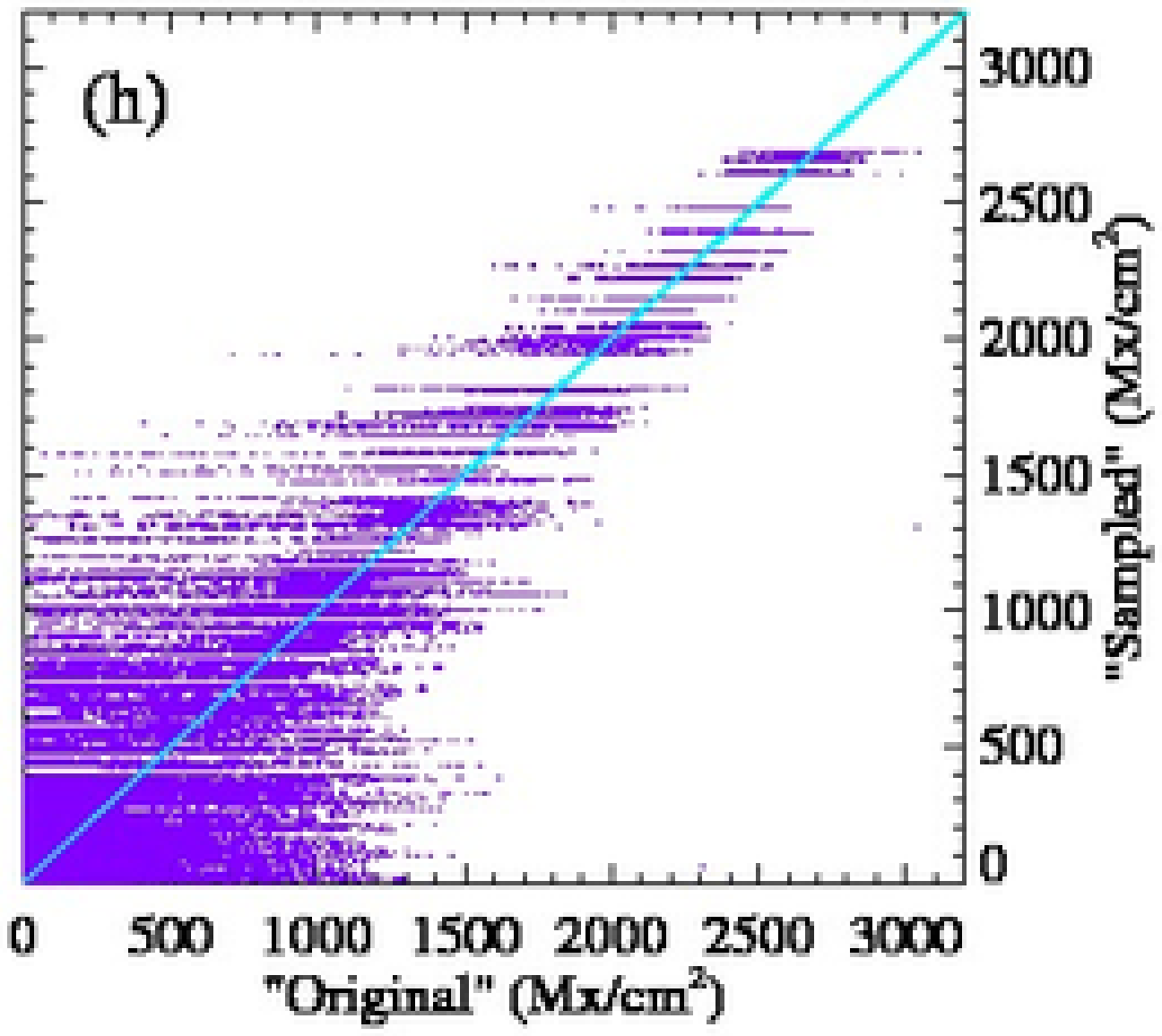}}
\vspace{0.25cm}
\caption{Follows Figure~\ref{fig:flowers_btot_scat} for (a)-(d), except
comparing the original {\it Hinode}/SP data with bin-factor 16 results. Figures
(e)-(h) follow the same format, but for the product $f\times \B$.}
\label{fig:hinode_btot_scat} 
\end{figure}

Scatter plots of the inverted manipulated spectra demonstrate the
general averaging which results with worsening spatial resolution
(Figure~\ref{fig:hinode_btot_scat}).  Of note in the {\it Hinode}/SP data,
compared to the synthetic case (Figure~\ref{fig:flowers_btot_scat}),
is the much greater spread in the original-resolution field strengths
compared to the binned results.  This behavior occurs
primarily in ``weak field'' or weak-polarization areas, where determining the
field strength and fill fraction independently is arguably problematic;
but that is not the case for all pixels.   The product
$f\times \B$ is also shown; the distributions do
change perceptibly (contrary to the synthetic case), with decreased scatter 
in weak-signal areas. (However, recall that only the ``instrument''
binning result is an independent inversion).  Primary contenders for the different 
behavior between $\B$ and $f \times \B$ here,
compared with the synthetic data, include the effects of photon noise
and the contention that the original-resolution
{\it Hinode}/SP data are unresolved to begin with.

\begin{figure}
\centerline{
\includegraphics[width=1.0\textwidth]{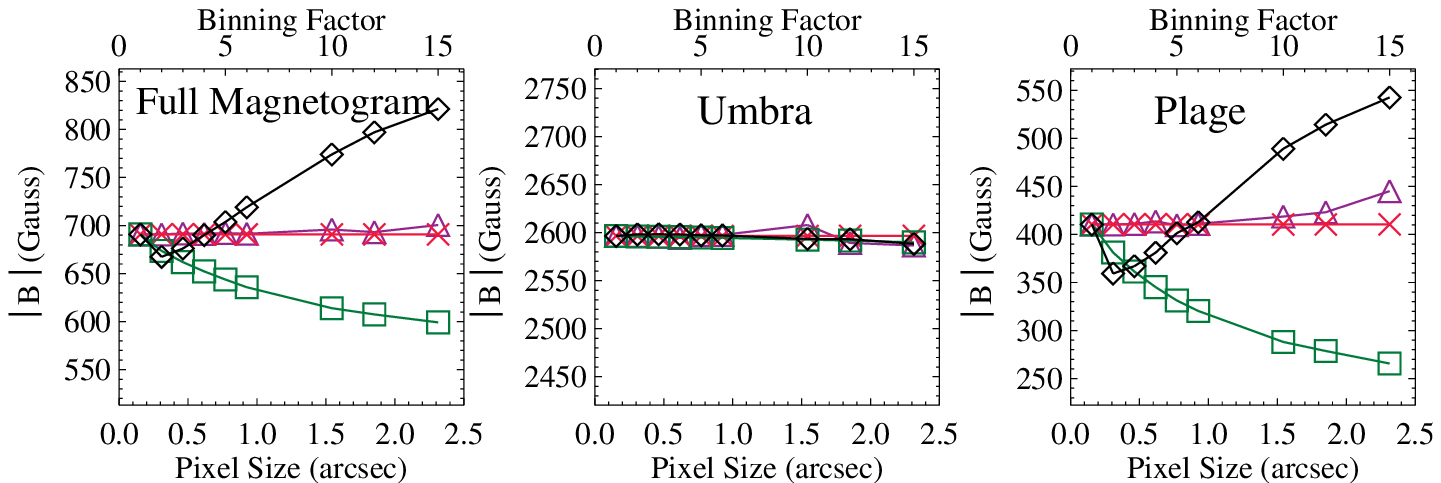}}
\vspace{-0.10cm}
\caption{\small Following Figure~\ref{fig:flowers_BB}, the 
the average field strength over the target area, $\frac{1}{N} \sum{\B}$ 
as a function of binning factor (top $x$-axis), for the four binning methods 
({\bf ``instrument'' $\Diamond$}, {\color{green}``bicubic'' $\Box$}, {\color{red} ``average'' $\times$}
and {\color{violet} ``sampled'' $\triangle$}),
focusing on three areas as indicated: the full magnetogram, an ``umbral''
area and a ``plage'' area, as depicted in Figure~\ref{fig:hinode}.  
With the original {\it Hinode}/SP scan resolution of $0.15\arcsec$,
the resulting pixel sizes are also indicated (bottom $x$-axis).  }
\label{fig:hinode_BB}
\end{figure}

Changes in the inferred magnetic field distribution in the observational
data show similar
trends with binning factor as was seen in the synthetic data.
Beginning with field strength (Figure~\ref{fig:hinode_BB}),
the umbral area shows little change, but the plage area is quite
sensitive to bin factor and to method used.  The full field of view
behaves closest to the plage.  

\begin{figure}
\centerline{\includegraphics[width=1.0\textwidth]{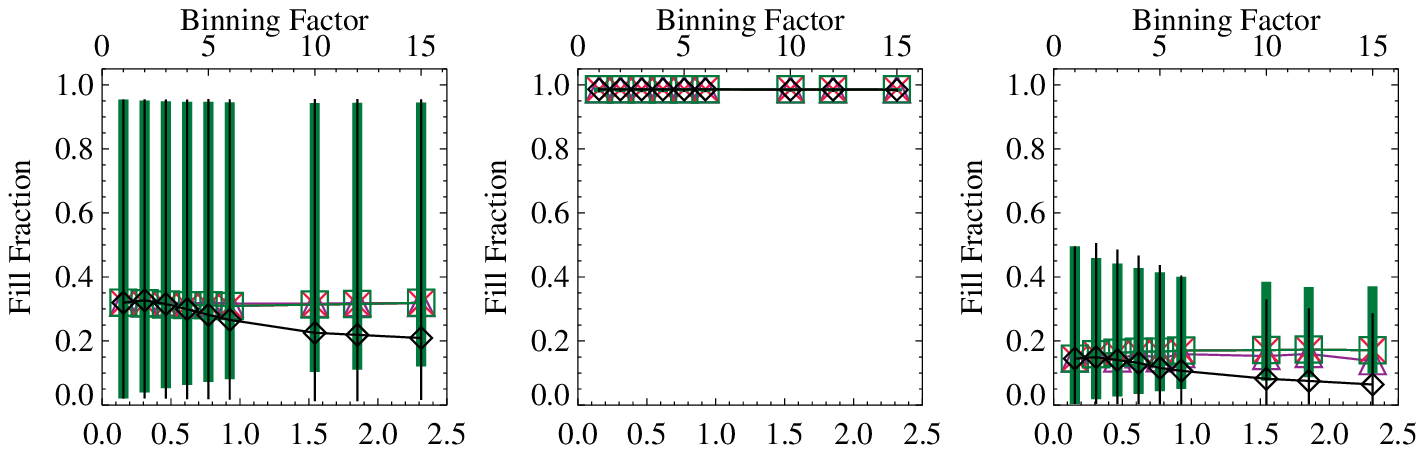}}
\vspace{-0.5cm}
\centerline{\includegraphics[width=1.0\textwidth]{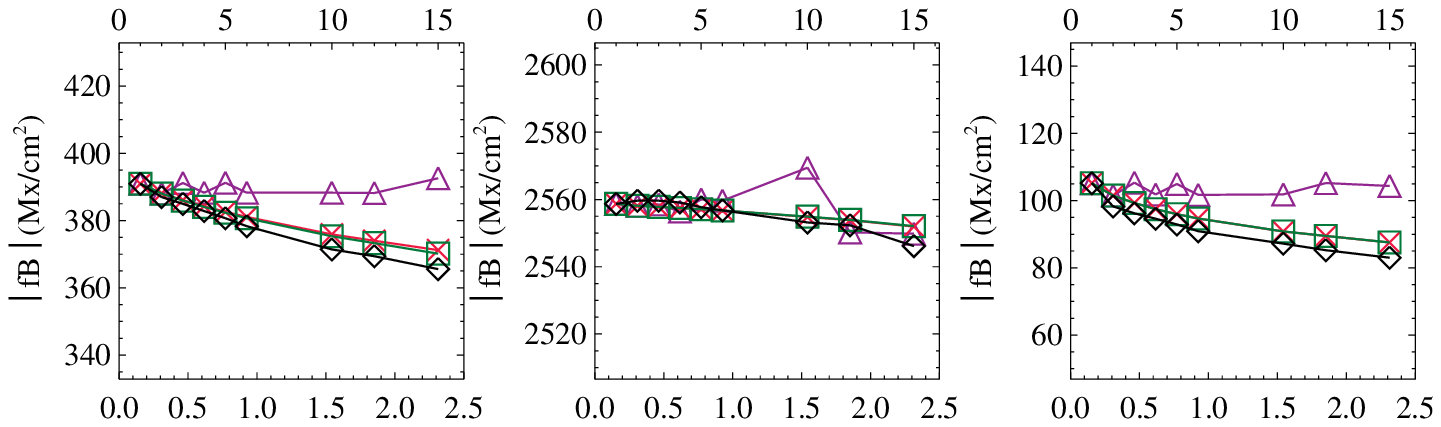}}
\vspace{-0.5cm}
\centerline{\includegraphics[width=1.0\textwidth]{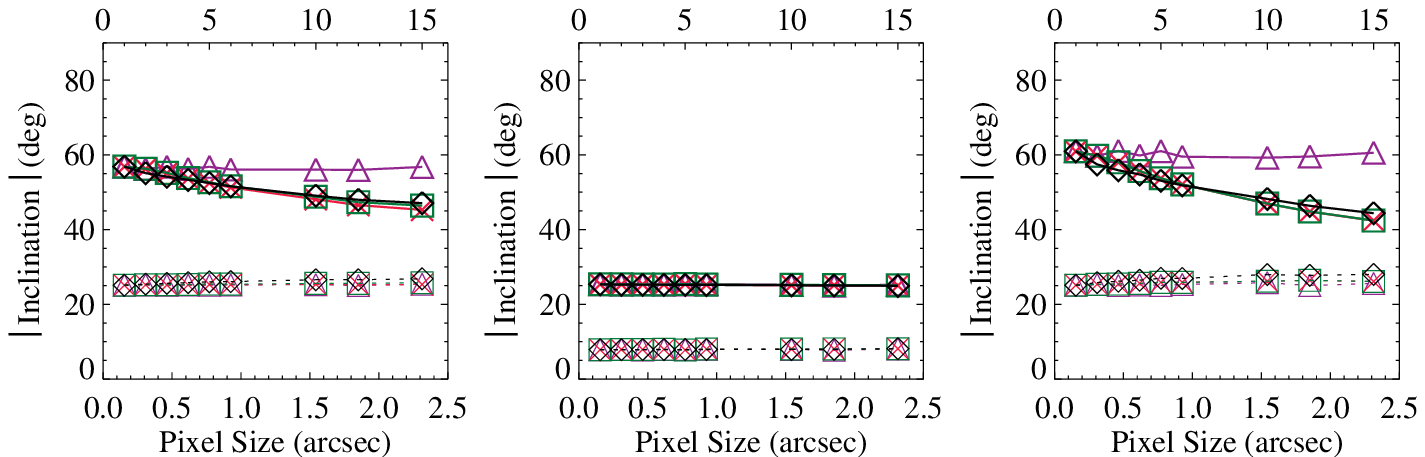}}
\vspace{-0.15cm}
\caption{\small Following the following the format of
Figure~\ref{fig:hinode_BB}, Top row: median (symbols) and 10th,~90th 
percentiles (displayed as ``error bars'') of inferred magnetic fill fraction as a function of
bin factor.  The three ``post-facto'' approaches consistently return
the same fill fraction as the original observations, as expected.
Middle row: the average product of the fill fraction and field
strength, $\frac{1}{N} \sum{f\B}$ as a function of binning factor.
Bottom row: variation of the average inclination angle with
binning factor (thick line-connected curves), $0^\circ$ indicates
(unsigned) fields directed along the line of sight, or pure $\Bl$,
and $90^\circ$ indicates field perpendicular to the line of sight
or pure $\Bt$ (here, $\gamma=\tan^{-1}(\Bt,|\Bl|)$).  Dot-connected
curves indicate the standard deviation of the angle distribution. }
\label{fig:hinode_fffBgamma} 
\end{figure}

The other inferred parameters examined here, the fill fraction,
product $f \times \B$, and instrument-frame inclination
(Figure~\ref{fig:hinode_fffBgamma}) confirm the general behavior
observed in the synthetic-data experiments.  The {\it Hinode}/SP data
start with a wide range of inferred fill fraction present, and a
median of less than 50\% at full resolution for the full field of view
(Figure~\ref{fig:hinode_fffBgamma}).  Again, the three post-facto binnings
do an averaging or sampling, hence the mean of the fill fraction
distribution stays the same although the range of values present
decreases with bin factor.  The ``instrument'' binning results in a
decreasing mean and tighter range as the spatial resolution degrades,
indicating that areas which were resolved become less so.

The product $f \times \B$ shows a systematic decrease, on average,
with worse spatial resolution -- except from the ``sampling'' approach,
which stays relatively constant.  The ``instrument'' approach displays
the most variation with resolution change, but the difference between
it and the other methods is not as dramatic compared to the experiment
with the model data.

The results for field inclination (Figure~\ref{fig:hinode_fffBgamma}),
show a distinct trend of the field becoming more aligned with the line of
sight with decreasing spatial resolution, especially in the plage areas.
In the umbra, there is effectively no change in the inclination angle
distribution.  The imperturbability of ``sampling'' against variations
in inclination was seen earlier, as well; again, the sampling should
represent the underlying field distribution (until the super-pixels are
themselves large and the resulting number of binned pixels available is
small), since it samples rather than averages.  We present the image-plane
inclination angle from the line of sight -- closely related to the direct
observables, but related to the physical inclination of the field to
the local normal by way of the observing angle.  Since $\mu=0.98$ for
these data, the difference between image-plane and the heliographic-plane
inclination from the local vertical direction is minimal.

\begin{figure}
\centerline{\includegraphics[width=1.0\textwidth]{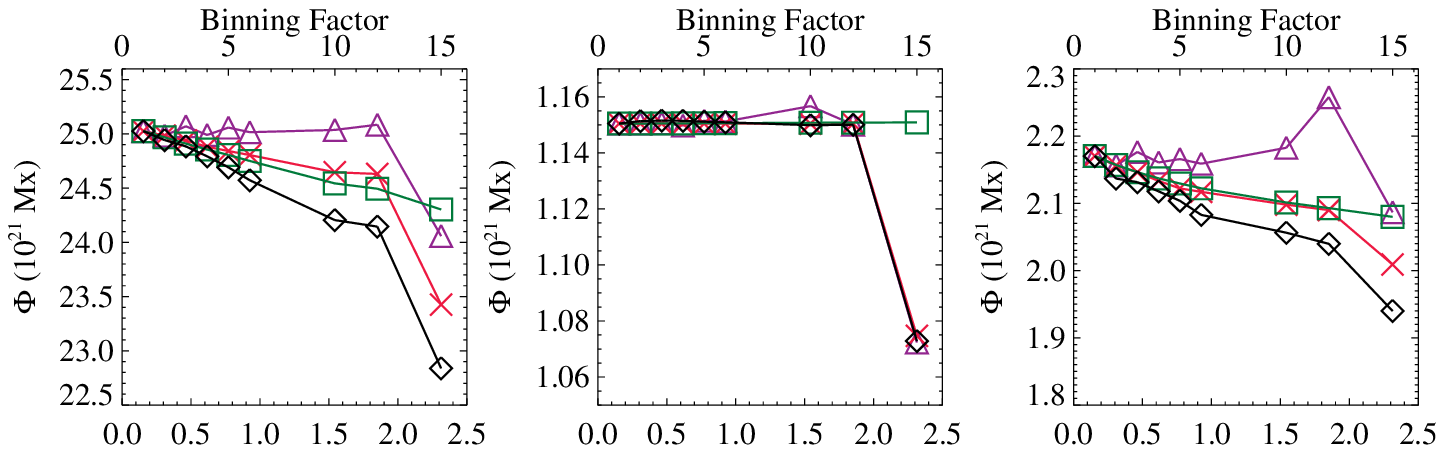}}
\vspace{-0.5cm}
\centerline{\includegraphics[width=1.0\textwidth]{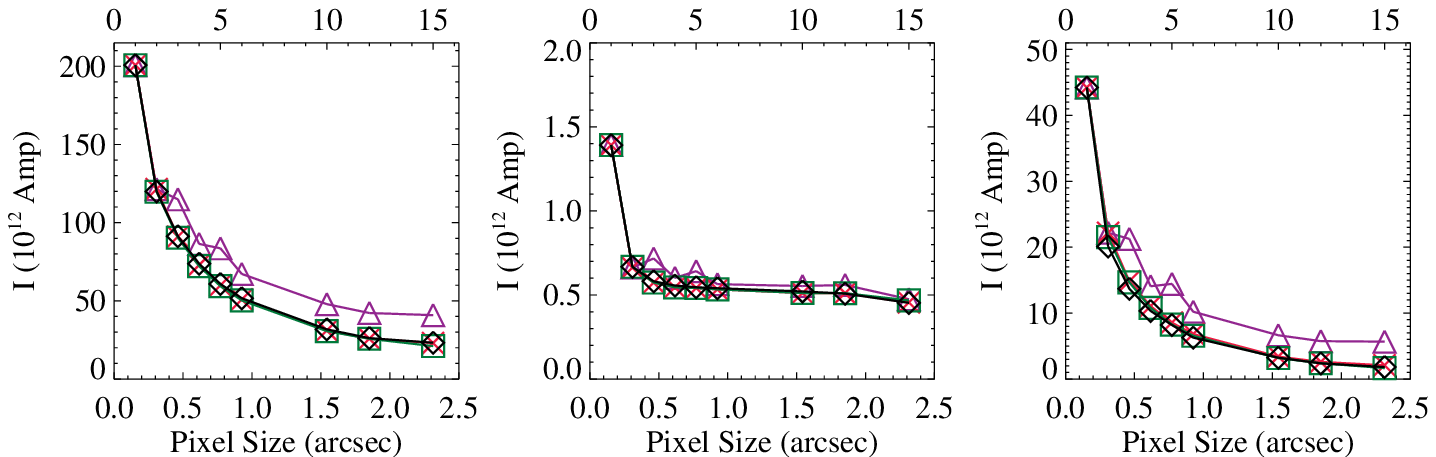}}
\vspace{-0.15cm}
\caption{\small Comparison of parameters often used for characterizing
active regions.  Top: Variation of the total unsigned magnetic flux 
$\Phi=\sum{f |\Bz| d{\rm A}}$
Bottom: of the total unsigned electric current $I=\sum{|\Jz| d{\rm A}}$.
For these plots, the $y$-axis is allowed to vary.}
\label{fig:hinode_flux}
\end{figure}

The total unsigned magnetic flux, $\Phi = \sum{f|\Bz| dA}$ behaves
essentially the same in the umbral areas of both synthetic and {\it Hinode}
data, varying little with resolution (except when there are arguably very
few points within that area of interest, see Figure~\ref{fig:hinode_flux}).
And again, the full field-of-view behavior is dictated by what kind of
structure dominates at highest resolution.
We see that the ``instrument'' spectral binning and subsequent inversion,
which is designed to mimic decreasing telescope size,
produces more of an effect than the ``post-facto'' binning approaches.

The total vertical electric current is often used to parametrize
an active region's stored magnetic energy (\inlinecite{params} and references therein).  Could this
characterization differ as a function of spatial resolution?  In the
{\it Hinode}/SP data, for all fields of view, there is a smooth decrease
of total current with decreasing spatial resolution. In addition, all
binning methods appear to act identically in this case.  The behavior
of the {\it Hinode}/SP data most resembles the synthetic ``plage'' beyond
bin-factor 10.  That is, the observational data, even at $0.15\arcsec$,
most closely resembles the area filled with unresolved multiple
small-scale magnetic centers.

Overall, the plage area observed with {\it Hinode}/SP produces the most
variations due to rebinning or degraded spatial resolution.  The umbral
area is least sensitive.  The sampling typically provides the most consistent
answer, but is also susceptible to the particular point sampled.
The effect of changing the instrumental resolution more closely
follows the results of the post-facto approaches as compared to the
trends in the simulation data. 
Assuming that the behavior of the full magnetogram is characterized
by the relative fraction of ``resolved'' or near-unity fill fraction
pixels within the field of view, it is clear that the {\it Hinode}/SP data
are dominated by non-unity fill fraction pixels and unresolved field
structure, even at the highest resolution.

\begin{figure}[t]
\centerline{\includegraphics[width=1.0\textwidth]{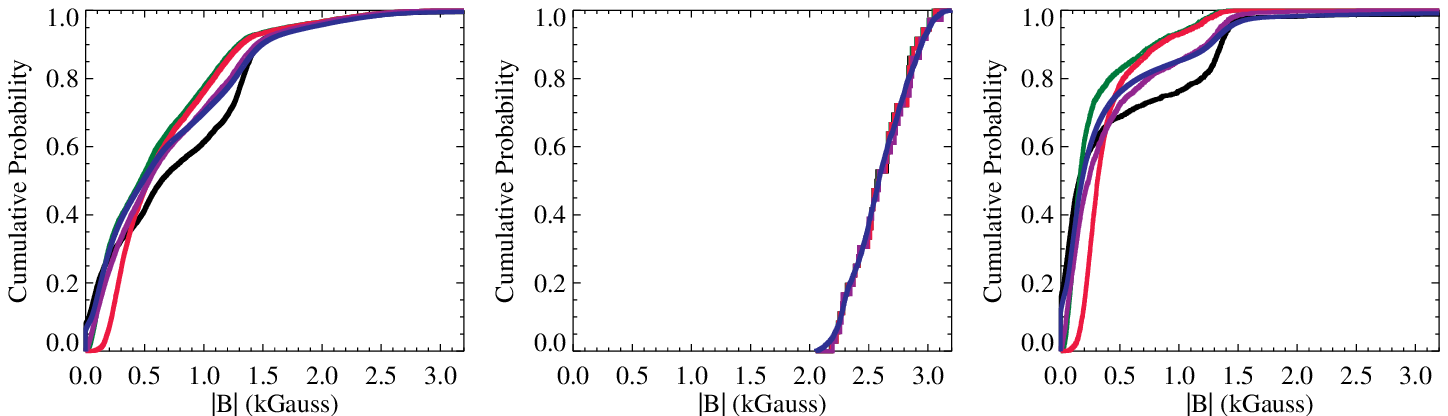}}
\vspace{-0.5cm}
\centerline{\includegraphics[width=1.0\textwidth]{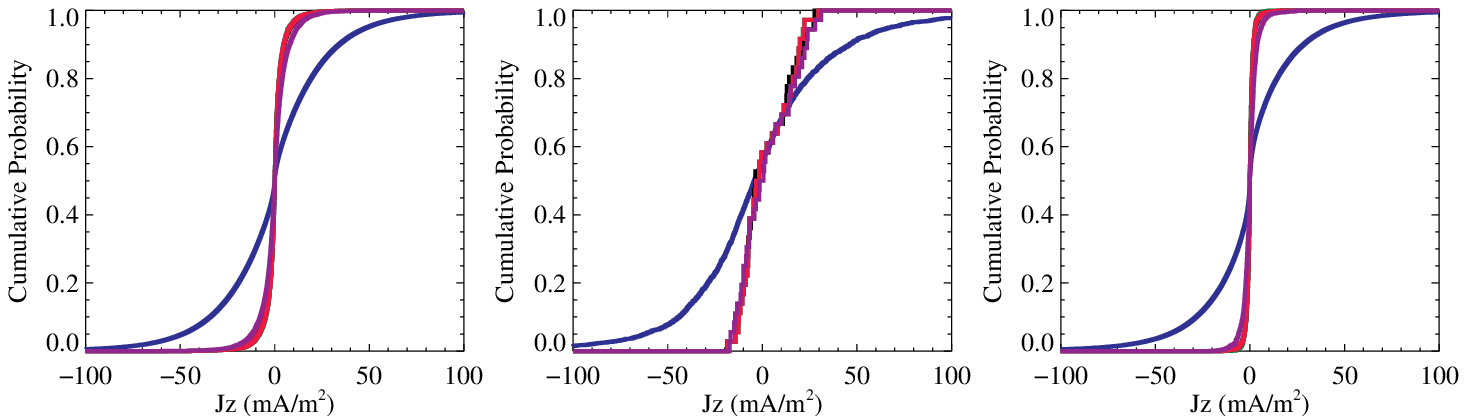}}
\caption{\small Cumulative probability distributions,
for the full-resolution data and the bin-10 results, for 
the three fields of view (entirety, ``umbra'', and ``plage'' areas).  
For each, CPD curves are plotted for: {\color{blue} original resolution},
{\bf instrument-binning}, {\color{green} bicubic}, {\color{red} average} and
{\color{violet} sampled} approaches.
Top: intrinsic field strength $\B$; Bottom: vertical electric
current density $\Jz$.}
\label{fig:hinode_cpd}
\end{figure}

The Kolmogorov-Smirnoff tests confirm statistically what is
described above.  The cumulative probability curves for field strength
(Figure~\ref{fig:hinode_cpd}), comparing the bin-10 results to the
original-resolution for both ``instrument'' and post-facto binnings
indicate distinct differences in the full field of view which is
reminiscent of the behavior in the plage area.  The umbral field
strength CPD looks almost identical to the umbral CPD for the synthetic
data (Figure~\ref{fig:flowers_cpd}).  The distribution of the vertical
current density (Figure~\ref{fig:hinode_cpd}) shows an almost exactly
opposite behavior than was observed in the synthetic data, in that the
original resolution indicates the presence of inferred vertical current
which has decreased in magnitude significantly at bin-factor 10.

\begin{figure}
\centerline{\includegraphics[width=1.0\textwidth]{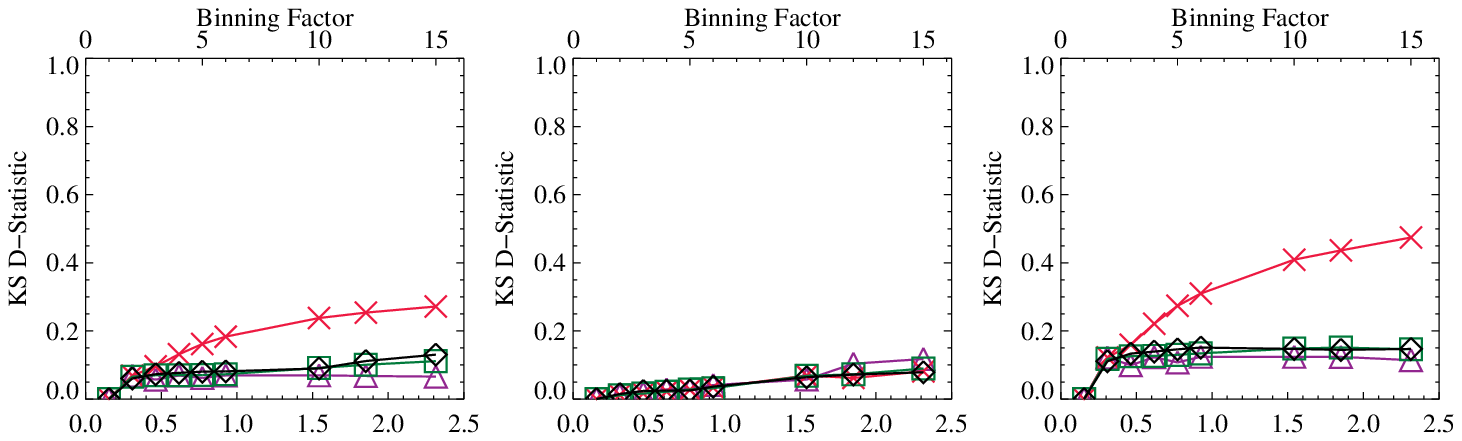}}
\vspace{-0.5cm}
\centerline{\includegraphics[width=1.0\textwidth]{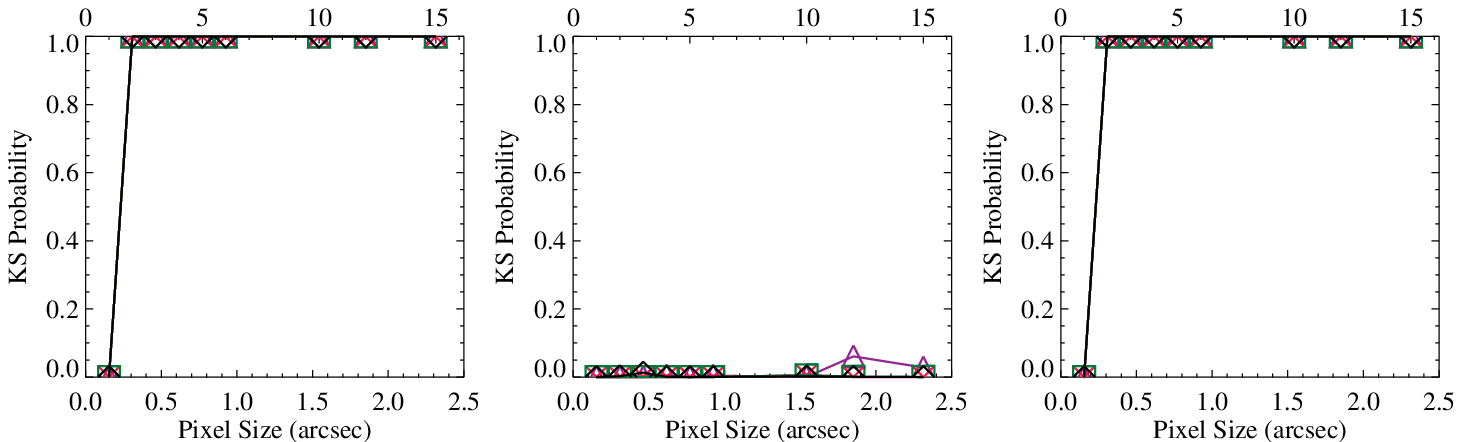}}
\caption{\small Summaries of the 
Kolmogorov-Smirnoff tests as a function of binning factor, for the
intrinsic field strength $\B$ over the three fields of view. 
Top: the ``D'' statistic; Bottom: the probability
that the two samples are {\it different} (see text).
Shown are curves for the original resolution {\it vs.} the ({\bf ``instrument'' $\Diamond$}, 
{\color{green}``bicubic'' $\Box$}, {\color{red} ``average'' $\times$}
and {\color{violet} ``sampled'' $\triangle$}) vertical current
distributions.}
\label{fig:hinode_ks_btot}
\end{figure}

The K-S statistics for field strength are more consistent across bin
factors (Figure~\ref{fig:hinode_ks_btot}) than in the synthetic data:
the D-statistic is slightly elevated but only varies dramatically with
the simple binning.  The K-S probability is unity for the plage area and the
full field of view for all bin factors, indicating that the samples are
not drawn from the same population.  On the contrary, it can be argued
that areas with consistent unity fill fraction statistically sample the
same population as the underlying field.

\begin{figure}
\centerline{\includegraphics[width=1.0\textwidth]{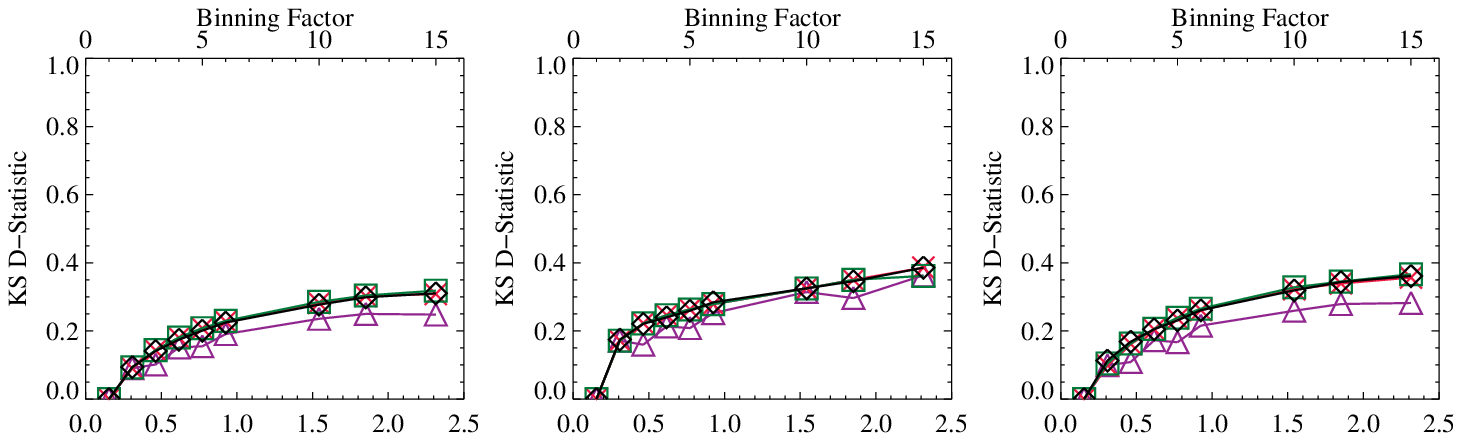}}
\vspace{-0.5cm}
\centerline{\includegraphics[width=1.0\textwidth]{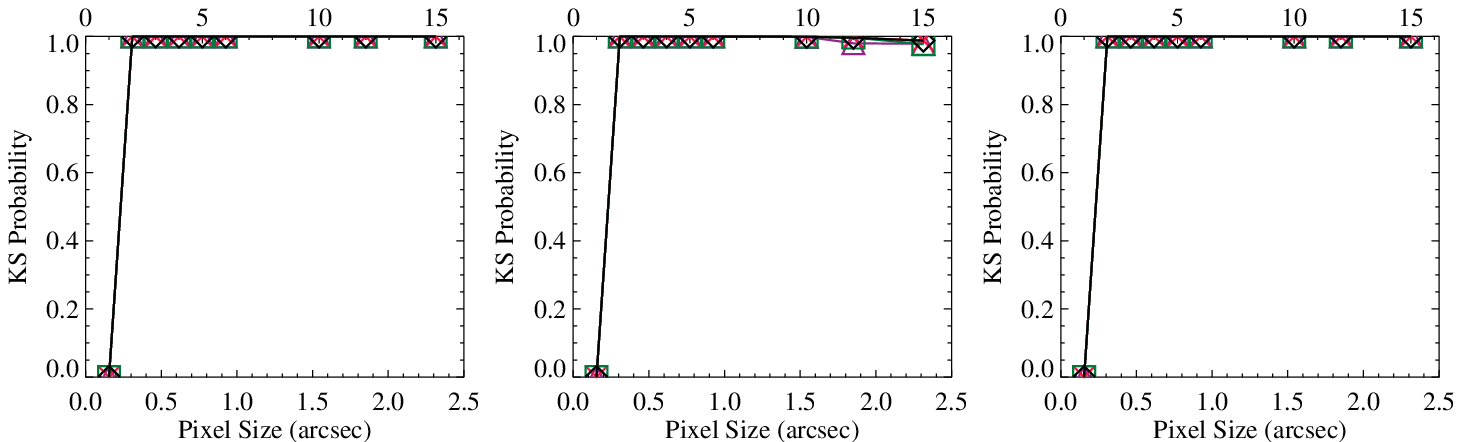}}
\caption{\small Summaries of the
Kolmogorov-Smirnoff tests as a function of binning factor, for the vertical electric
current density $\Jz$, following Figure~\ref{fig:hinode_ks_btot}.}
\label{fig:hinode_ks_jz}
\end{figure}

On the other hand, the vertical current density is affected at all spatial
resolutions (Figure~\ref{fig:hinode_ks_jz}). From a statistical
point of view the results from lower resolution data do not represent
the underlying distribution of the highest spatial resolution, even
in the unity-fill-fraction umbral area.  One may simply conclude that
the actual distribution of vertical current in the solar photosphere
is unknown and unknowable without absolutely full resolution
everywhere in question.

\section{Summary and Conclusions}

We outline a manner by which to manipulate Stokes polarization spectra
in order to mimic the effects of instrumental spatial resolution to the
simplest order.  Through the use of a synthetic magnetic field construct that is
both fully resolved and contains small-scale structures, we apply
this method to a range of degradations.  We find (not surprisingly)
that it is the highly structured areas which are most sensitive
to the effects of instrumental optical spatial resolution.  

The analysis indicates (also not surprisingly) that even the {\it Hinode}/SP ``normal
scan'' spectropolarimetric data at $0.15\arcsec$ spatial sampling
are unresolved.  Recalling this, plus the fact that we could only bin
up to a factor of 16 before completely decimating the number of pixels
needed for analysis, the patterns shown by field parameters with
degrading resolution are remarkably similar to those found using the
synthetic data.  We thus confirm the appropriateness of the findings
from these synthetic-data as valid for helping interpret the observational data.

Statistical tests confirm that whether by instrumental spectral mixing
or post-facto methods, worsening spatial resolution results in a map
of the vector field which does not reproduce the underlying magnetic
structure, except in select areas where the returned magnetic filling
factor in the binned data is still unity.  Where the returned fill
factor is less than unity, worsening spatial resolution leads to an
average image-plane inclination angle more aligned to the line of sight,
an increasing average field strength which couples with the decreasing
average fill fraction to present a decreasing total magnetic flux.  The behavior
of further-derived parameters that rely on spatial derivatives is less
straightforward, but may impart a non-zero current density and inferred
``twist'' where there in fact are none.  The pessimistic interpretation
of these results is that without the highest spatial resolution, the
underlying field is unrecoverable.  The optimistic interpretation is
that by making use of the inferred magnetic fill fraction for inverted
spectropolarimetric data, it is possible to tell where these effects
will be most dramatic, and where they will be least impactful.

The influence of spatial resolution on the instrument-plane inclination angle
implies that the impacts on physically-interpreted
variables in the (coordinate-transformed) heliographic plane will vary with 
observing angle.  This also has implications for our understanding
of the large-scale ``weak-field'' areas from instruments of limited
spatial resolution: in this context, the assumption that the photospheric field is dominantly radial
\cite{wangsheeley92,arge_etal_02} must be re-examined\footnote{As are the results 
that they may be predominantly horizontal, see \inlinecite{borrero_kobel_11}}.

Details and caveats to the above statements are important to mention.
There is no model of instrumental scattered light applied to (or
subsequently corrected for) the synthetic data; in parallel, the
{\it Hinode}/SP data are inverted using a common but simple treatment of
computing a scattered light profile, rather than a more sophisticated
local approach which has been demonstrated to better recover low-signal
areas \cite{orozco_etal_2007}.  While the details will differ had
we used the latter, the approach taken here is consistent, and hence
still illustrative.  Effects as drastic as shown here are generated in
the synthetic data without Doppler velocities or field gradients along
the line of sight, whereas both are expected for observational data.
Yet in the ``simple is OK'' defense, key behavior patterns are seen
clearly in the {\it Hinode}/SP data. 

We also ask how well instrumental resolution can be represented by ``post-facto''
manipulation of the vector field map.  Tests of three different methods
show that, again, in highly structured underlying areas,
these methods result in very different outcomes than expected from 
differences in aperture size.   Simply put, there are only special cases
where ``binning down'' a magnetogram will adequately mimic the differences
between different instrumental spatial resolutions, and generally the 
``instrument'' binning results in the largest differences from the underlying field. 

This exercise of comparing the results when one simulates ``worse
spatial resolution'' by different means is illuminating, and
demonstrates that method matters according to the goal of the study
in question.  Three basic categories are: comparisons/calibrations between
instruments, utilizing data from different instruments as part of
an analysis for which data from a single instrument falls short (due
to limited field of view, capability, availability, etc.), and interpreting
numerical simulation results in the context of observations.  

Regarding the first category, we note that while a
few instrument-comparison studies perform spatial
averaging on the polarization signals for comparison
\cite{wangetal92,ivm2}\footnote{\inlinecite{ivm2} performs a
near-simultaneous comparison between the IVM and the ASP, contrary to
the note in \inlinecite{bergerlites2002}, Section~1.1.}, the majority such
studies published thus far use some form of ``post-facto'' averaging and
binning applied to the magnetogram from the higher-resolution instrument
\cite{bergerlites2002,bergerlites2003,mdi_mwso_2005,demidovetal2008,wangd_etal_2009a,wangd_etal_2009b}.
It is clear that an ``instrument''-binning should be the preferred
method, since all post-facto approaches result in a different (and
typically smaller) variation with binning factor than expected from
optical resolution.

An addendum to this category is using synthetic data for tests
of algorithms through ``hare \& hound'' exercises, where the evaluation
depends crucially that the synthetic data mimic the behavior of
those real data eventually slated for analysis.  As such, including
the gross effects of the instrument or observing method chosen
\cite{ambigworkshop2,orozco_etal_2007}
is needed in order to not arrive at incorrect conclusions.

In the second category, if the goal is to preserve the underlying
character of the vector magnetic field {\it and} the region in question
has a high average filling fraction, then post-facto binning can be
employed with some confidence.  However, as was shown with the vertical
current density, while the magnetic field distribution and character
may be preserved, quantities that are derived from the field must be
viewed with less confidence.  This is a very restrictive set of caveats,
but the most well-defended position according to this study.

The third category acknowledges the great strides in simulations
of solar magnetic structure, and the approach of validating
them quantitatively using comparisons to observed structures
\cite{pores,abbett07,orozco_etal_2007,sheminova09}.  It is insufficient
to rebin or apply a blurring function directly to a simulation's
well-resolved output for comparisons to the solar observations.  We 
reiterate that, due to these results, at the
very least a simple modeling and manipulation of emergent spectra 
is required for even qualitative comparisons between simulations and observations of
the magnetic field distribution.

In this context, we come back to Table~\ref{table:mdi_sp_comp1} (see
Appendix~\ref{sec:mdi_hinode_comp}, Table~\ref{table:mdi_sp_comp2} and
Figure~\ref{fig:mdi_hinode_comp}).  The minimal impact of the post-facto
``congrid'' approach on the {\it Hinode}/SP fast-scan map ``Flux''=$\sum{|\Bl|}$
result is consistent with what we have shown here.  The MDI Level 1.8.1
data used in section~\S\ref{sec:realdata1} and in \inlinecite{nlfff3}
present a systematic offset from the Level 1.8.2 calibration (which
became available December 2008, and decreased the $\Bl$ magnitudes
by $\approx8--9$\% in the location of AR~10953\footnote{See
{\tt http://soi.stanford.edu/magnetic/Lev1.8/ for details}}). 
When variations in field of view,
calibration, and especially spatial resolution are accounted for
according to the findings of this paper (details can be found in
Appendix~\ref{sec:mdi_hinode_comp}), there still exists an offset
between the results from MDI and {\it Hinode}/SP that is larger than the
quoted uncertainties, but may still be attributable to remaining
differences in the lines' formation heights and inversion methods.

Finally, from this investigation, it is still unclear what the solar
magnetic field structure actually {\it is}, especially for areas with
fine-scale structure.  This is not a new concept \cite{mismas2},
but reinforced here through a simple, yet thorough demonstration.
We show that our ignorance is especially true for quantities
derived from the vector field maps which rely on spatial derivatives
\cite{parker96,ambigworkshop2}.  Are vector magnetic field maps useless?
Definitely not!  Comparisons between data of active regions obtained with
consistent instrumentation and spatial resolution do detect differences
amongst the structures that must, somehow, be related to the inherent
magnetic structure, especially as manifest in the release of stored
magnetic energy ({\it e.g.}~\inlinecite{dfa3} and references therein).  But in the context of
measuring and interpreting the state and behavior of the solar plasma,
conclusions that are drawn must do so in the context of the limitations
of the data employed.

\begin{acks}

KDL first acknowledges Dr. Richard C. Canfield, who introduced
her to spectropolarimetry and the interpretation of vector magnetic
field maps (at $6\arcsec$ resolution!).  We also appreciate the supportive and helpful
comments from the referee.
This work was made possible by the models and instruments developed under the following
respective funding sources:
NASA contracts 
NNH05CC75C, 
NNH09CE60C
and NNH09CF22C, 
the NWRA subcontract from the Smithsonian
Astrophysical Observatory under NASA NNM07AB07C,
and the NWRA subcontract from Stanford University NASA Grant NAS5-02139
for SDO/HMI commissioning and pipeline code implementation.
We thank Dr. Bruce Lites at NCAR/HAO
for reformatter code and updates to the HAO inversion code.
We also sincerely thank Mr. Eric Wagner
for understanding ``scientist code'' and making things actually run.
{\it Hinode} is a Japanese mission developed and launched
by ISAS/JAXA, collaborating with NAOJ as a domestic partner, NASA and
STFC (UK) as international partners. Scientific operation of the {\it
Hinode} mission is conducted by the {\it Hinode} science team
organized at ISAS/JAXA, consisting of scientists from 
institutes in the partner countries. Support for the post-launch
operation is provided by JAXA and NAOJ (Japan), STFC (U.K.), NASA,
ESA, and NSC (Norway). MDI data are provided by the SOHO/MDI consortium. 
SOHO is a project of international cooperation between ESA and NASA.

\end{acks}

\appendix

\section{Constructing Representative Instrument-Binned Spectra When Only Demodulated Spectra Are
Available}\label{sec:photonnoise}

For an instrument like {\it Hinode}/SP, the demodulation from six states is
performed onboard the spacecraft, so only the four demodulated states are
available.  Since the demodulated states do not contain all the information of
the original states, we discuss here the impact of this loss of information
on the noise level of the reconstructed states. 

Assuming that each of the six polarization states actually observed at a given
wavelength, $(I \pm P)(\lambda)$, is
drawn from a Poisson distribution, the expectation value of each
distribution is given by $\langle I \pm P \rangle \equiv p_\pm^\lambda$, where
$P$ can be any of $Q$, $U$ or $V$.  Since each of these is a Poisson
distribution, the variance of each is equal to the expectation value.

The demodulated states actual available are given by
\begin{eqnarray}
P(\lambda) &\equiv& [(I + P)(\lambda) - (I - P)(\lambda)]/2 
\\
I(\lambda) &\equiv& [(I + Q)(\lambda) + (I - Q)(\lambda) + (I + U)(\lambda) 
+ (I - U)(\lambda) 
\nonumber \\ 
&& + (I + V)(\lambda) + (I - V)(\lambda)]/6.
\end{eqnarray}
(Henceforth, the wavelength dependence is assumed for clarity).
Working specifically with $I\pm Q$ as an example, since each modulated state
will have similar behavior, the reconstructed modulated states are
\begin{eqnarray}
(I \pm Q)_{\rm R} &=& I \pm Q 
\nonumber \\ 
&=& {1 \over 6} \bigg[(I + Q) + (I - Q)
+ (I + U) + (I - U) 
\nonumber \\ 
&& + (I + V) + (I - V) \bigg ] \pm {1 \over 2} 
\bigg [(I + Q) - (I - Q) \bigg ] 
\nonumber \\ 
&=& {2 \over 3} (I \pm Q) - {1 \over 3} (I \mp Q)
\nonumber \\ 
&& + {1 \over 6} \bigg [(I + U) + (I - U) + (I + V) 
+ (I - V) \bigg ] 
\end{eqnarray}
which has an expectation value of 
\begin{eqnarray}
\langle (I \pm Q)_{\rm R}(\lambda) \rangle &=& {2 q_\pm^\lambda \over 3} - {q_\mp^\lambda \over 3} 
+ {u_+^\lambda + u_-^\lambda + v_+^\lambda + v_-^\lambda \over 6}.
\end{eqnarray}
and a variance of
\begin{eqnarray}
{\rm var} (I \pm Q)_{\rm R}(\lambda) &=& {4 q_\pm^\lambda \over 9} + {q_\mp^\lambda \over 9} 
+ {u_+^\lambda + u_-^\lambda + v_+^\lambda + v_-^\lambda \over 36},
\end{eqnarray}
whereas the expectation value and the variance of the actual state is simply
$q_\pm^\lambda$.  In the continuum (or anywhere the polarization is low), this
reduces to
\begin{eqnarray}
{\rm var} (I \pm Q)_{\rm R}^{\rm c} &=& {4 q_\pm^{\rm c} \over 9} + {q_\mp^{\rm c} \over 9} 
+ {u_+^{\rm c} + u_-^{\rm c} + v_+^{\rm c} + v_-^{\rm c} \over 36} 
\nonumber \\ 
&\approx& {2 \over 3} p^{\rm c}. 
\end{eqnarray}
Thus the variance in the reconstructed modulated states, at least in areas of
weak polarization, is smaller than the variance in the original states. Further, 
since each reconstructed state is the sum of six Poisson variables, rather than 
being a single Poisson variable (at a given wavelength), the distribution of 
the noise will also differ.  

\section{Comparing MDI and {\it Hinode}/SP Line-of-Sight ``Flux''}
\label{sec:mdi_hinode_comp}

As presented in this manuscript, instruments with different resolutions
will provide quantitatively different descriptions of the solar
magnetic field.  We began the study with a provocative ``why are
these the same, and why are those different?'' example.  In detail,
of course, there is more to this than simply the spatial resolution of
two different instruments.  The MDI data used for \inlinecite{nlfff3}
were from the level 1.8.1 calibration, the {\it Hinode}/SP data were provided
by B.W.~Lites, with ostensibly the same inversion that was used here
for the ``instrument''-binning exercise (although 
probably with slightly different implementation), but which
also included a remapping to square pixels using an unknown method.
Not only the spatial sampling but the field of view differs between
the {\it Hinode}/SP scans of 18:35 and 22:30 UT, as one can see by closely
examining Figures~\ref{fig:10953blos}~and~\ref{fig:hinode}.

\begin{figure}[t]
\includegraphics[width=1.0\textwidth]{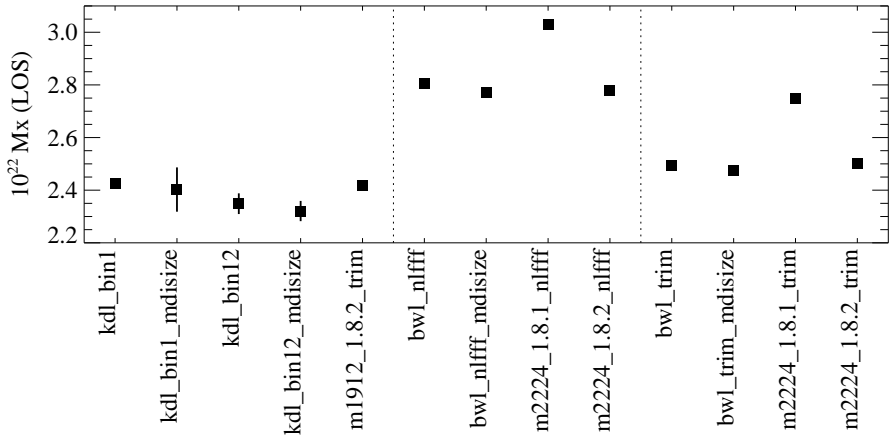}
\caption{\small The total unsigned ``flux'' $\Phi_{\rm los} = \sum{|\Bl| dA}$
for NOAA AR 10953 on 30 April 2007 from various sources and 
methods of spatial resolution modeling.  See Table~\ref{table:mdi_sp_comp2} and text 
for descriptions of tags and the three groups indicated by vertical lines.
Formal error bars are included for each point.}
\label{fig:mdi_hinode_comp}
\end{figure}

Here we demonstrate just how sensitive comparisons can be to the details
of calibration, inversion, and very slight variations in the physical
area sampled.  Table~\ref{table:mdi_sp_comp2} summarizes
differences in the data sources and processing to obtain maps of AR
10953 on 30 April 2007, and Figure~\ref{fig:mdi_hinode_comp} shows the
variation in the inferred $\Phi_{\rm los} = \sum{|\Bl| dA}$ for each.
A full propagation of uncertainties was performed, for MDI following
\inlinecite{hagenaar_2001}, for {\it Hinode}/SP data using the uncertainties
returned from inversions and propagated for $\Bl$.  For most points,
the uncertainty is smaller than the plotting symbol.

The entries are combined into three rough groups.  The first is based
on the 18:35UT {\it Hinode}/SP ``normal'' scan used for most of this paper,
and the total $\sum{|\Bl| dA}$ based on it is deemed the reference.
Entries include the results from post-facto ``sampling'' to match the
MDI resolution, and an ``instrument'' bin-12 to get close to the MDI
resolution, with an additional sampling from that to match it exactly
as indicated.  These are compared to the MDI level 1.8.2 dataset closest
in time.

The second and third groups are based on the 22:30 {\it Hinode}/SP ``fast'' scan
(see \S~\ref{sec:realdata1}), a post-facto ``sampled'' map based on it,
and comparisons to the closest-time MDI level-1.8.1 and level-1.8.2 data.
The difference between these two groups is whether the full 22:30
{\it Hinode}/SP is used or whether all are trimmed to match the (slightly)
smaller field of view of the 18:35UT {\it Hinode}/SP scan.  The difference
is not so slight.

Clearly, the binning approaches behave as described in the text, however
those effects are insignificant as compared to even small discrepancies in
the field of view and calibration.  And evolution: comparing datasets
which are as consistent as possible but separated by time, we see the
active region increasing its total magnetic signal during this period
(see \opencite{okamoto_etal_08}).

The answer to the small puzzle presented in Section~\ref{sec:realdata1}
is that in fact the level-1.8.1 MDI calibration produced systematically
larger $\Phi_{\rm los}$ results than could otherwise be explained
by spatial resolution issues; this is mostly accounted for by the
recalibrated MDI level-1.8.2 data, as these examples show.  And as we have demonstrated,
post-facto manipulation of a magnetogram as in Section~\ref{sec:realdata1}
does not generally reproduce the differences in instrumental spatial
resolution.  There exists still a small offset such that the level-1.8.2
data return a $\Phi_{\rm los} = \sum{|\Bl| dA}$ greater by a few percent
than expected from the quoted uncertainties when the best possible
match is compared (Table~\ref{table:mdi_sp_comp2}, ``kdl\_bin12'' and
``kdl\_bin12\_mdisize'' vs. ``m2224\_1.8.2\_trim''.).  We acknowledge this
is a single example, and invoke spectral-line properties and inversion
method differences as probable contributors.

\begin{table}[h]
\begin{center}
\caption{AR 10953, 30 April 2007, Total $\sum{|\Bl| dA}$ Details.}
\label{table:mdi_sp_comp2}
\begin{tabular}{lccc} \hline
Label & Data source/Time & Details/Area &  Difference (\%) \\ \hline
kdl\_bin1 & {\it Hinode}/SP 18:35 & ``instrument'' bin-1 &  -- --\\
 & ``normal'' scan &  & \\
kdl\_bin1\_mdisize & '' '' & ``kdl\_bin1''+{\tt congrid} $\rightarrow1.98\arcsec$ & -1 \\
kdl\_bin12 & '' '' & ``instrument'' bin-12 & -3 \\
kdl\_bin12\_mdisize & '' '' & ``kdl\_bin12''+{\tt congrid} $\rightarrow1.98\arcsec$ &-4 \\
m1912\_1.8.2\_trim & MDI 19:12UT\footnote[1]{} & [387:446,429:525]\footnote[2]{} & -0.4 \\
 & Level 1.8.2  & & \\ \hline
bwl\_nlfff & {\it Hinode}/SP 22:30  & Inversion by B.W.Lites for & +15 \\
 & ``fast'' map & \inlinecite{nlfff3}. &  \\
 &  & Remapped to square pixels & \\
bwl\_nlfff\_mdisize & '' '' & ``bwl\_nlfff''+{\tt congrid}$\rightarrow1.98\arcsec$ & +14 \\
m2224\_1.8.1\_nlfff & MDI 22:24UT\footnote[3]{} &  [385:460,429:509] & +25 \\ 
   & Level 1.8.1 & & \\ 
m2224\_1.8.2\_nlfff &  Level 1.8.2 & '' '' & +14 \\  \hline
bwl\_trim & {\it Hinode}/SP 22:30 & Trimmed in $x$-dir to & +3 \\
& ``fast'' map  & match {\it Hinode}/SP 18:35 &  \\
bwl\_trim\_mdisize & '' '' & ``bwl\_trim''+{\tt congrid}$\rightarrow1.98\arcsec$ & +2 \\
m2224\_1.8.1\_trim & MDI 22:24UT\footnote[3]{} & [400:461,428:509] &+13 \\
 & Level 1.8.1  & &  \\
m2224\_1.8.2\_trim & Level 1.8.2 & '' '' & +3 \\ \hline
\end{tabular}
\end{center}
\vspace{-0.8cm}
\flushleft{
1: {\tt fd\_M\_96m\_01d.5232.0012.fits}\\
2: Indexing starts at 0 \\
3: {\tt fd\_M\_96m\_01d.5232.0014.fits} \\
}
\end{table}

\bibliographystyle{spr-mp-sola-cnd}
\bibliography{resolution}

\end{article} 

\end{document}